\documentclass[11pt,fleqn]{article}

\usepackage{amsmath,amssymb,amsthm}
\usepackage[margin=1in]{geometry}
\usepackage{xcolor}
\usepackage{url}
\usepackage{array}
\usepackage{comment}
\usepackage{thm-restate}
\usepackage{booktabs}
\usepackage{tikz}
\usepackage{enumitem}
\usepackage[hypertexnames=false,bookmarksnumbered=true,final]{hyperref}
\usepackage{algorithm,algpseudocode}
\usepackage[capitalize,sort]{cleveref}

\def\colorschemesepia{sepia}
\def\colorschemedark{dark}
\def\colorschemelight{light}

\ifx\colorscheme\undefined
\let\colorscheme\colorschemelight
\fi

\ifx\colorscheme\colorschemelight
\colorlet{textColor}{black}
\colorlet{bgColor}{white}
\fi

\ifx\colorscheme\colorschemesepia
\definecolor{textColor}{HTML}{433423}
\definecolor{bgColor}{HTML}{fbf0da}
\fi

\ifx\colorscheme\colorschemedark
\definecolor{textColor}{HTML}{bdc1c6}
\definecolor{bgColor}{HTML}{202124}
\definecolor{textBlue}{HTML}{8ab4f8}
\definecolor{textRed}{HTML}{f9968b}
\definecolor{textGreen}{HTML}{81e681}
\definecolor{textPurple}{HTML}{c58af9}
\else
\colorlet{textBlue}{blue!50!black}
\colorlet{textRed}{red!50!black}
\colorlet{textGreen}{green!50!black}
\definecolor{textPurple}{HTML}{681da8}
\fi

\ifx\colorscheme\colorschemelight\else
\pagecolor{bgColor}
\color{textColor}
\fi

\hypersetup{colorlinks,linkcolor=textRed,citecolor=textRed,urlcolor=textBlue}

\newtheorem{theorem}{Theorem}
\newtheorem{definition}{Definition}
\newtheorem{example}{Example}
\newtheorem{corollary}{Corollary}
\newtheorem{claim}{Claim}
\newtheorem{lemma}{Lemma}

\newcommand*{\defeq}{:=}
\newcommand*{\Th}{^{\textrm{th}}}

\newcommand*{\chat}{\widehat{c}}
\newcommand*{\ghat}{\widehat{g}}
\newcommand*{\hhat}{\widehat{h}}
\newcommand*{\shat}{\widehat{s}}
\newcommand*{\vhat}{\widehat{v}}
\newcommand*{\Ghat}{\widehat{G}}
\newcommand*{\Hhat}{\widehat{H}}

\newcommand*{\Xhat}{\widehat{X}}

\newcommand*{\vord}{\vhat}
\newcommand*{\ordInst}{\widehat{\mathcal{I}}}
\newcommand*{\Xord}{\Xhat}
\newcommand*{\cord}{\chat}
\newcommand*{\gord}{\ghat}
\newcommand*{\hord}{\hhat}
\newcommand*{\Gord}{\Ghat}
\newcommand*{\Hord}{\Hhat}

\DeclareMathOperator{\MMS}{MMS}

\DeclareMathOperator{\ord}{\mathtt{Order}}
\DeclareMathOperator{\celim}{\mathtt{EnvyCycleElimination}}
\DeclareMathOperator{\ttcelim}{\mathtt{TTEnvyCycleElimination}}
\DeclareMathOperator{\psequence}{\mathtt{PickingSequence}}
\DeclareMathOperator{\pick}{\mathtt{Pick}}

\DeclareMathOperator{\oneLess}{oneLess}

\DeclareMathOperator{\resolve}{resolve}

\newcommand{\instance}{\mathcal{I}}
\newcommand{\mms}{{\normalfont\textsc{MMS}}}

\newcommand{\mxs}{{\normalfont\textsc{MXS}}}
\newcommand{\eefx}{{\normalfont\textsc{EEFX}}}
\newcommand{\pls}{{\normalfont\textsc{PLS}}}
\newcommand{\efx}{{\normalfont\textsc{EFX}}}
\newcommand{\eefxZero}{{\normalfont\textsc{EEFX}}$_0$}
\newcommand{\efxZero}{{\normalfont\textsc{EFX}}$_0$}
\newcommand{\prop}{{\normalfont\textsc{PROP}}}
\newcommand{\propOne}{{\normalfont\textsc{PROP1}}}
\newcommand{\propx}{{\normalfont\textsc{PROPX}}}
\newcommand{\ps}{{\normalfont\textsc{PS}}}
\newcommand{\ef}{{\normalfont\textsc{EF}}}

\newcommand{\sefx}{{\normalfont\textsc{Efx}}}

\newcommand{\naturals}{\mathbb{N}}

\newcommand{\np}{{\normalfont\textsc{NP}}}

\newcommand{\bpartition}{{\normalfont\textsc{BalancedPartition}}}
\newcommand{\partition}{{\normalfont\textsc{Partition}}}

\newcommand{\set}[1]{\{#1\}}

\newcommand{\agents}{[n]}

\newcommand*\circled[1]{\tikz[baseline=(char.base)]{\node[shape=circle,draw,inner sep=2pt] (char) {#1};}}

\def\versionOpre{opre}

\def\VERSION{default}
\newenvironment{noQedProof}[1][Proof]{\hskip\labelsep{\it #1}.\enskip }{}
\let\citet\cite
\let\citep\cite
\DeclareMathOperator*{\argmin}{argmin}
\DeclareMathOperator*{\argmax}{argmax}
\def\Halmos{\mbox{\quad$\square$}}%

\allowdisplaybreaks

\title{New Fairness Concepts for Allocating Indivisible Items}
\author{
Ioannis Caragiannis%
\thanks{Department of Computer Science, Aarhus University, Denmark}
\\ \texttt{\small iannis@cs.au.dk}
\and
Jugal Garg%
\thanks{Department of Industrial \& Enterprise Engineering, University of Illinois at Urbana-Champaign, USA}
\\ \texttt{\small jugal@illinois.edu}
\and
Nidhi Rathi%
\thanks{Max-Planck-Institute for Informatics, Saarbrücken, SIC, Germany}
\\ \texttt{\small nrathi@mpi-inf.mpg.de}
\and
Eklavya Sharma\footnotemark[2]
\\ \texttt{\small eklavya2@illinois.edu}
\and
Giovanna Varricchio%
\thanks{Department of Mathematics and Computer Science, University of Calabria, Italy}
\\ \texttt{\small giovanna.varricchio@unical.it}
}
\date{\empty}

\begin{document}

\maketitle

\newcommand{\ABSTRACT}[1]{#1}
\begin{abstract}
\ABSTRACT{%
For the fundamental problem of \emph{fairly} dividing a set of indivisible items among agents, \emph{envy-freeness up to any item} (EFX) and \emph{maximin fairness} (MMS) are arguably the most compelling fairness concepts proposed until now.  Unfortunately, despite significant efforts over the past few years, whether EFX allocations always exist is still an enigmatic open problem, let alone their efficient computation. Furthermore, today we know that MMS allocations are not always guaranteed to exist. These facts weaken the usefulness of both EFX and MMS, albeit their appealing conceptual characteristics.

We propose two alternative fairness concepts---called \emph{epistemic EFX} (EEFX) and \emph{minimum EFX share fairness} (MXS)---inspired by EFX and MMS. For both, we explore their relationships to well-studied fairness notions and, more importantly, prove that EEFX and MXS allocations always exist and can be computed efficiently for additive valuations. Our results justify that the new fairness concepts can be excellent alternatives to EFX and MMS.
}

\end{abstract}

\section{Introduction}
\label{sec:intro}

Fair division is a popular research area, with origins in antiquity and important applications~\citep{brams1996fair,moulin2004fair}. In a setting that has received much attention in the last fifteen years, a set of indivisible items is to be distributed {\em fairly} among agents. But what does ``fairly'' mean? There is no single answer here and different ways of interpreting ``fairly'' have been considered in the literature.

The first interpretation is {\em comparative}. To evaluate an allocation as fair, each agent compares the bundle of items allocated to her with bundles of the remaining items. The well-known notion of {\em envy-freeness}~\citep{foley1966resource} is a representative fairness concept defined this way, according to which an allocation is fair if each agent prefers the bundle allocated to her to the bundle allocated to any other agent. Typically, agents have valuation functions that allow them to evaluate or compare bundles of items.

The second interpretation is {\em in absolute terms}. In this category, each agent defines a {\em threshold} based on her view of the allocation instance and evaluates as fair those allocations in which she gets a bundle of value that exceeds this threshold. {\em Proportionality}~\citep{dubins1961cut,steinhaus1948problem} is the representative fairness concept here. Each agent's threshold is simply her total value for all items divided by the number of agents. Then, an allocation is proportional if each agent gets a bundle of items for which her value exceeds her proportionality threshold.

Unfortunately, a seemingly attractive fairness concept may not be useful in the setting of indivisible items we consider. Even though it is undeniably hard to argue that an envy-free allocation is unfair, envy-freeness has at least two drawbacks. First, it may not be feasible to achieve. Consider the problem faced by a library where three computer science books (the items) are to be given to two CS students (the agents). Whichever student gets less than two books will be envious of the other. Second, finding an envy-free allocation may be a computationally challenging problem. It is not hard to see that deciding whether an envy-free allocation among two agents with identical valuations for the items exists is at least as hard as deciding the paradigmatic \np-hard \partition\ problem; e.g., see~\citet{bouveret2016characterizing}. Proportionality suffers from the same {\em infeasibility} and {\em high computational complexity} drawbacks as well.

Several relaxations of envy-freeness proposed in the literature aim to circumvent this issue. The first one, {\em envy-freeness up to some item} (\ef1), introduced by~\citet{budish2011combinatorial}, requires that each agent prefers her own bundle to the bundle of any other agent, after removing some item from the latter. \ef1 addresses the two above-mentioned issues in an ideal way. \ef1 allocations always exist and can be computed efficiently~\citep{lipton2004approximately}. Unfortunately, it seems that \ef1 has moved way too far and has lost the fairness properties of envy-freeness. For example, assume that the two students in our example have a high value for one of the three books. Then, the allocation that gives one low-value book to one of them is \ef1. This student is indifferent to the bundle of the other student after removing the high-value book from it.

Intuitively, it is clear that a more fair allocation would give the two low-value books to one of the students and the high-value book to the other. This is what motivates the definition of {\em envy-freeness up to any item} (\efx), introduced by \citet{CKMPSW19}. An allocation is \efx\ if each agent prefers her own bundle to the bundle of any other agent, after removing {\em any} item from the latter. From the fairness point of view, \efx\ is almost as appealing as envy-freeness. However, despite significant efforts by researchers in the last decade~\citep{plaut2020almost,caragiannis2019envy,chaudhury2020efx,amanatidis2021maximum,chaudhury2021little,mahara2021extension,berger2021almost,ChaudhuryGMMM21,AkramiACGMM23}, it is still unknown whether \efx\ is always feasible for instances with more than three agents. Furthermore, even for the case of three agents where the existence of \efx\ allocations is guaranteed, the computational complexity of the known methods is immense~\citep{chaudhury2020efx,mahara2021extension}.

Among the relaxations of proportionality, the one that has received the lion's share of attention uses the so-called {\em maximin fair share} (\mms), i.e., the maximum value an agent can attain in any allocation where she is assigned her least preferred bundle, as threshold. Surprisingly, \citet{kurokawa2018fair} proved that \mms\ allocations may not always exist. Since then, research has focused on computing allocations that approximate \mms; e.g., see~\citet{amanatidis2017approximation,barman2020approximation,GhodsiHSSY21,garg2020improved,FST21,AkramiG24}. These \mms-approximations are less appealing as fairness concepts. An excellent recent survey by \citet{survey2022} discusses the above fairness concepts and many more.

\subsection{Our Conceptual and Technical Contribution}

The discussion above suggests that the Holy Grail of research in fair division with indivisible items is to define a concept that is (1) intuitively, as close to fairness as envy-freeness and proportionality, (2) always feasible, and (3) efficiently computable. In this paper, we present two such concepts, inspired by \efx\ and \mms. Our first one, called {\em epistemic envy-freeness up to any item} (\eefx), is comparative and adapts the concepts of {\em epistemic envy-freeness} defined by~\citet{ABCGL18}. An allocation is \eefx\ if, for every agent, it is possible to shuffle the items in the remaining bundles so that she becomes ``\efx-satisfied''. An example follows.

\begin{example}\label{example:firstExample}
Consider a fair division instance with three agents 1, 2, and 3, having additive valuations, and eight items $a$, $b$, \ldots, $h$, with the valuations depicted in the next table. Each row has the values of an agent for the items. The circles indicate the allocation $(\{a,d,e\},\{b,c,f,g\},\{h\})$.
\begin{table}[ht]
{\em \begin{tabular}{cccccccc}
\toprule  $a$ & $b$ & $c$ & $d$ & $e$ & $f$ & $g$ & $h$
\\ \midrule
  \circled{$40$} &  $2$ &  $2$ & \circled{$15$} & \circled{$15$} &  $2$ &  $2$ &  $2$
\\  $4$ &  \circled{$10$} & \circled{$10$} &  $2$ &  $2$ & \circled{$25$} & \circled{$25$} &  $2$
\\ $20$ &  $4$ &  $4$ & $7$ & $7$ &  $10$ &  $4$ & \circled{$24$}
\\ \bottomrule
\end{tabular}}
\end{table}

\noindent This allocation is \eefx. Agents 1 and 2 are not envious of any other agent. For agent 3, by reshuffling the items $a, \ldots, g$ in the bundles of agents 1 and 2, we can get the allocation $(\{a,b,c\},\{d,e,f,g\},\{h\})$, in which agent 3 is \efx-satisfied (note that the definition of \eefx\ requires that the reshuffling does not affect the bundle of agent 3). Her value of $24$ for her bundle $\{h\}$ is at least as high as her value for bundle $\{a,b,c\}$ after removing her least-valued item $b$ and for bundle $\{d,e,f,g\}$ after removing her least-valued item $g$.
\end{example}

\eefx\ is particularly relevant in environments (with privacy restrictions or of very large scale) in which each agent knows the allocation instance (i.e., she is aware of all items and her valuations for them, as well as of the number of agents) but her knowledge of the allocation is limited to the contents of her bundle. Then, the agent is as {\em optimistic} as possible in the evaluation of her bundle; she compares it with each bundle in the allocation of the remaining items that would be best possible for her. In contrast to \eefx, the concepts of envy-freeness, \ef1, and \efx\ are {\em knowledge-sensitive} according to the definitions of~\citet{ABCGL18}.

Our second fairness concept interprets fairness in absolute terms by defining the {\em minimum \efx\ share} (\mxs) as a threshold for each agent. The minimum \efx\ share of an agent is the minimum value she has among all allocations in which she is \efx-satisfied.

\begin{example}\label{example:secondExample}
We compute the minimum \efx\ share of agent 2 in \cref{example:firstExample} and show that it is equal to $25$. Notice that the agent is \efx-satisfied in allocation $(\{a,g\},\{f\},\{b,c,d,e,h\})$. Indeed, her value is $25$, which is also her value for bundle $\{a,g\}$ after removing her least-valued item $a$. Her value for the other bundle $\{b,c,d,e,h\}$ after removing her least-valued item $d$ is only $24$. We can see that agent 2 is not \efx-satisfied in any allocation that gives her a value of $24$. In this case, agent 2 gets her value of $24$ by items $b$ and $c$ and at most two other items among $a$, $d$, $e$, and $h$. The items $f$ and $g$ should be in the first and third bundle, respectively. These bundles should also have at least two items among $a$, $d$, $e$, and $h$. Then, agent 2 will be envious of some of these two bundles after removing her least-valued item. Thus, the minimum \efx\ share of agent 2 is $25$.
\end{example}

\mxs\ allocations are similar in spirit to proportional and \mms\ allocations. An allocation is \mxs\ if each agent gets a bundle of value at least her minimum \efx\ share. We define our two new fairness concepts formally in \cref{sec:prelims}.

We present the following technical results about \eefx\ and \mxs. First, in \cref{sec:relation}, we explore their relation to \mms\ as well as to {\em proportionality up to one item} (\prop1), another relaxation of proportionality~\citep{CFS17}. We show that every \mms\ allocation is \eefx, every \eefx\ allocation is \mxs, and every \mxs\ allocation is \prop1. This chain of implications puts the new fairness concepts in the spectrum of existing fairness criteria and extends previous taxonomies by \citet{bouveret2016characterizing} and \citet{ABCGL18}. To the best of our knowledge, the fact that every \mms\ allocation is also \prop1\ was not known before.

Our main result is a polynomial-time algorithm that computes an \eefx\ (and, hence, \mxs) allocation in any fair division instance. Our analysis exploits several key ideas from the fair division literature, such as the concept of {\em ordered instances}, which are typically used in the design of approximation algorithms for \mms\ since the work of \citet{bouveret2016characterizing}, the {\em envy cycle elimination algorithm} of \citet{lipton2004approximately}, and the fact that applying this algorithm on ordered instances results in \efx\ allocations, observed independently by~\citet{plaut2020almost} and \citet{barman2020approximation}. Our algorithm is essentially identical to an algorithm used by \citet{barman2020approximation} to compute $2/3$-\mms\ allocations. Due to our different fairness objective, our analysis is considerably different and rather simpler. In addition to \eefx, we get \mxs\ and $2/3$-\mms\ as bonus properties for the allocations computed by our algorithm. These results appear in \cref{sec:eefx}.

Rather surprisingly, in \Cref{sec:hardness}, we show that computing the minimum \efx\ share is an \np-hard problem. This suggests that any efficient algorithm for computing \mxs\ allocations cannot make explicit use of the minimum \efx\ share.

In \Cref{sec:chores}, we turn our attention to the case of chores, that is, where items are negatively valued. In this scenario, we prove that our result on the existence of \eefx\ carries over. In particular, with a careful adaptation of the algorithm presented in \citet{barman2020approximation} for computing an approximate \mms\ allocation of chores, we simultaneously obtain \eefx\ and $4/3$-\mms. Unfortunately, the implications $\mms \Rightarrow \eefx$ and $\mxs \Rightarrow \propOne$ hold for goods but not for chores. However, we get a new interesting implication, $\eefx \Rightarrow \propx$ for chores. Clearly, by definition, every \eefx\ allocation is \mxs.

In \cref{sec:cancelable}, we show that \eefx\ allocations also exist for \emph{cancelable} valuations, a generalization of additive valuations. We conclude with a discussion on extensions of our new fairness concepts and many open problems in \cref{sec:discussion}.

\section{Definitions and Notation}
\label{sec:prelims}

Throughout the paper, we use $[k]$ to denote the set $\{1, 2, ..., k\}$ for a non-negative integer $k$.

A fair division instance $\instance$ consists of the set of $n$ \emph{agents} and $m$ (indivisible) \emph{items}. We use positive integers to identify both the agents and the items and denote their sets as $[n]$ and $[m]$, respectively.
Each agent $i$ has a \emph{valuation function} $v_i: 2^{[m]} \to \mathbb{R}$, i.e., for any set $S$ of items, $v_i(S)$ denotes how valuable $S$ is to agent $i$. We assume $v_i(\emptyset) = 0$ for all $i \in [n]$.
A valuation function $v_i$ is said to be \emph{additive} if $v_i(S) = \sum_{g \in S} v_i(\{g\})$ for all $S \subseteq [m]$.
For simplicity, we often write $v_i(g)$ instead of $v_i(\{g\})$.
When $v_i(S_1) \le v_i(S_2)$, for all $S_1 \subseteq S_2 \subseteq [m]$,
we say that the items are \emph{goods}.
Unless stated otherwise, we assume that the items are goods and agents' valuation functions are additive.

An \emph{allocation} $X = (X_1, X_2, \ldots, X_n)$ of the items to the agents is a partition of $[m]$ into $n$ bundles $X_1, X_2, \ldots, X_n$, with $X_i$ being the bundle allocated to agent $i$. Formally, $X_i \cap X_j = \emptyset$ for all $i \neq j$ and $\cup_{i=1}^n{X_i}=[m]$. We often use $\Gamma$ to refer to the set of all allocations of a given instance.

\subsection{Notions of Fairness}
\label{sec:prelims:notions}

{\em Envy-freeness up to any item} (EFX) is among the most compelling fairness concepts in the literature.

\begin{definition}[Envy-freeness up to any item (EFX)]
\label{defn:efx}
An agent $i$ is \emph{EFX-satisfied} by allocation $X = (X_1, \ldots, X_n)$ if for any other agent $k$,
and any item $g \in X_k$ such that $v_i(g) > 0$, it holds that $v_i(X_i) \ge v_i(X_k \setminus \set{g})$.
The allocation $X$ is \emph{EFX} if every agent is EFX-satisfied by $X$.
(We refer the reader to Appendix~\ref{sec:efx-defn} for a discussion on our choice of including the restriction $v_i(g) > 0$ in \cref{defn:efx}.)
\end{definition}

We are ready to define our first new fairness concept.

\begin{definition}[Epistemic \efx\ (\eefx) and \eefx\ certificates]
For a fair division instance, an allocation $X = (X_1, X_2, \ldots, X_n)$ is called {\em epistemic \efx\ (\eefx)}
if for every agent $i \in \agents$, there exists an allocation $Y = (Y_1, Y_2, \ldots, Y_n)$ such that
$Y_i = X_i$ and agent $i$ is \efx-satisfied with $Y$.
We refer to such an allocation $Y$ as an \emph{\eefx\ certificate} of agent $i$ for bundle $X_i$.
\end{definition}

In other words, we can say that an allocation $X$ is epistemic \efx\ if there exists an \emph{\eefx\ certificate} for every agent with respect to her bundle in $X$.
Note that, an \efx\ allocation trivially serves as an \eefx\ certificate for every agent,
and, thus, an \efx\ allocation is \eefx\ as well.
It can be easily seen that the opposite is true for two agents, but not for three or more agents as Example \ref{example:firstExample} indicates.

Our second new fairness concept is similar in spirit to the well-known \mms\ fairness property.

\begin{definition}[Maximin share (\mms)]
For a fair division instance, the \emph{maximin share} of an agent $i \in [n]$, denoted as $\MMS_i$, is defined as follows
\[ \MMS_i := \max_{Z=(Z_1, \ldots, Z_n) \in \Gamma}{\min_{j \in [n]}{v_i(Z_j)}} \, . \]
Moreover, we say that an allocation is \mms\ if every agent receives a bundle of value that is at least as high as her maximin share.
\end{definition}

Analogously, we define the {\em minimum EFX share} (\mxs) and {\em \mxs\ allocations}.
For a fair division instance, we use $\sefx_i$ to denote the collection of all allocations where
agent $i$ is $\efx$-satisfied. Formally,
\[ \sefx_i := \left\{\begin{array}{l}
    Z=(Z_1, \ldots, Z_n) \in \Gamma:
    \\ \displaystyle \quad v_i(Z_i) \geq \max_{j \in \agents} \max_{\substack{g\in Z_j:\\ v_i(g) > 0}}
        v_i(Z_j \setminus \set{g})
\end{array}\right\}. \]
Note that when the inner maximum is taken over an empty set, then it implies that agent $i$ values every item in bundle $Z_j$ at $0$, and hence the stated condition is trivially satisfied. We now define an agent's \emph{minimum \efx\ share} as the least value she derives from any allocation where she is $\efx$-satisfied.

\begin{definition}[Minimum \efx\ share, \mxs\ allocations]
For a fair division instance, we define the \emph{minimum \efx\ share} of agent $i$ as $\mxs_i := \min_{Z \in \sefx_i}{v_i(Z_i)}$.
Moreover, we say that $Z = (Z_1, \ldots , Z_n)$ is an \mxs\ allocation if $v_i(Z_i) \geq \mxs_i$ for every agent $i \in [n]$.
\end{definition}

One may wonder why we do not define a maximum \efx\ share fairness concept using the (similar to \mms$_i$) threshold of $\max_{Z\in \sefx_i}{\min_{j\in [n]}{v_i(Z_j)}}$ for agent $i$. Interestingly, we can show that this gives just an alternative definition for \mms.

We conclude this section with the definition of \emph{proportionality up to some item} (\propOne),
a variation of the well-known concept of proportionality.
\begin{definition}[Proportionality up to one item (PROP1)]
\label{defn:prop1}
For a fair division instance, we define agent $i$'s \emph{proportionality threshold} as
$\ps_i := v_i([m]) / n$. An allocation $X = (X_1, \ldots, X_n)$ is \emph{proportional up to one item} ($\propOne$) if
$\max_{g \in [m] \setminus X_i} v_i(X_i \cup \{g\}) \ge \ps_i$ for every agent $i \in [n]$.
\end{definition}

It is well-known -- e.g., see \cite{bouveret2016characterizing} -- that in every fair division instance, we have $\mms_i \leq \ps_i$ if $v_i$ is additive.

\subsection{Ordering}
\label{sec:prelims:order}

Let $i$ be an agent in a fair division instance $\instance$.
Let $(\ell_1, \ell_2, \ldots, \ell_m)$ be an ordering of the items such that $v_i(\ell_1) \ge v_i(\ell_2) \ge \ldots \ge v_i(\ell_m)$ (break ties by picking lower-indexed items first, i.e., $v_i(\ell_t) = v_i(\ell_{t+1}) \Rightarrow \ell_t < \ell_{t+1}$). Intuitively, $\ell_t$ is agent $i$'s $t\Th$ favorite item. For agent $i$, the \emph{rank} of an item $j$, denoted by $r_i(j)$, is the number $t$ such that $j = \ell_t$. For a set $S$ of items, define $r_i(S) \defeq \{r_i(j): j \in S\}$.

The \emph{ordered valuation function} of agent $i$, denoted by $\vord_i$, is defined as $\vord_i(T) \defeq v_i(\{\ell_t: t \in T\})$, for $T\subseteq [m]$. For any instance $\instance$, let $\ordInst \defeq \ord(\instance)$ be a fair division instance having the same set of goods and agents as $\instance$, but each agent $i$ has valuation function $\vord_i$ instead of $v_i$.

\begin{claim}
For every set $S$ of items and every agent $i$, $v_i(S) = \vord_i(r_i(S))$.
\end{claim}

A fair division instance is called \emph{ordered}, w.r.t.\ the natural ordering $1,\ldots, m$, if $v_i = \vord_i$ for each agent $i$.
In particular, for each agent $i$, this implies $v_i(\{1\}) \ge v_i(\{2\}) \ge \ldots \ge v_i(\{m\})$, and $r_i$ is the identity function.

\section{Relations to Other Fairness Concepts}
\label{sec:relation}

In this section, we establish interesting connections between our proposed fairness concepts of \eefx\ and \mxs\ with previously well-studied notions of fairness in the literature, summarized in the following chain of implications:
\[ \mms \Rightarrow \eefx \Rightarrow \mxs \Rightarrow \propOne \, . \]

The connection between \eefx\ and \mxs\ allocations follows easily by the definitions.

\begin{theorem}[$\eefx \Rightarrow \mxs$] \label{thm:eefx_mxs}
An \eefx\ allocation in a fair division instance is also \mxs.
\end{theorem}
\begin{proof}
Let $X=(X_1, \ldots, X_n)$ denote an $\eefx$ allocation in a fair division instance. Fixing an agent $i \in [n]$, we will prove that $v_i(X_i)$ is at least as high as her minimum \efx\ share. By definition, since $X$ is an \eefx\ allocation, there must exist an \eefx\ certificate $Y=(Y_1, \ldots, Y_n)$ for agent $i$ such that $Y_i=X_i$ and $Y\in \sefx_i$. Therefore, we can write
\[ v_i(Y_i)\geq \min\limits_{Z \in \sefx_i}v_i(Z_i) = \mxs_i. \]
Since, the bundles $X_i$ and $Y_i$ are identical, we obtain $v_i(X_i) \geq \mxs_i$, thereby completing the proof.
\end{proof}

The opposite implication, however, is not true (even for 2 agents having identical valuations)
as the following example shows.

\begin{example}[$\mxs \nRightarrow \eefx$]
\label{ex:eefx-nimpl-mxs}
Consider a fair division instance with 2 agents having identical additive valuations $v$ over 7 goods.
The first 2 goods have a value of $4$ each, and the remaining goods have a value of 1 each.

Agent 1 is EFX-satisfied by the allocation $([7] \setminus [2], [2])$, so $\mxs_1 \le 5$.
Agent 2 is EFX-satisfied by the allocation $([2], [7] \setminus [2])$, so $\mxs_2 \le 5$.
The allocation $X \defeq (\{1, 7\}, \{2, 3, 4, 5, 6\})$ is MXS, since each agent gets a bundle of value at least 5.
However, agent 1 not EFX-satisfied by $X$, since her value for agent 2's bundle is 7 after removing good 6.
Hence, $X$ is not an EEFX allocation.
\end{example}

We now present a non-trivial connection where we prove that any maximin share allocation necessarily admits an \eefx-certificate for every agent.

\begin{theorem}[$\mms \Rightarrow \eefx$] \label{thm:mms_eefx}
An \mms\ allocation in a fair division instance is also \eefx.
\end{theorem}
\begin{proof}
Consider a fair division instance. We begin with a definition.
For an allocation $Z = (Z_1, \ldots, Z_n)$ and an agent $i \in [n]$, we denote by $s^{Z,i}$ the vector containing the values $\{v_i(Z_j): j \in [n] \setminus \{i\}\}$ sorted in non-decreasing order. Hence, $s^{Z,i} \defeq \langle s^{Z,i}_1, s^{Z,i}_2, \ldots, s^{Z,i}_{n-1}\rangle$ satisfies $s^{Z,i}_t \le s^{Z,i}_{t+1}$ for all $t \in [n-2]$.

Let $X = (X_1, \ldots, X_n)$ be an MMS allocation for the fair division instance and let $i \in [n]$ be an arbitrary agent. Let $Y = (Y_1, \ldots, Y_n)$ be an allocation with $Y_i = X_i$ so that $s^{Y,i}$ is lexicographically minimum.
We will show that $Y$ is an EEFX certificate for agent $i$ and bundle $X_i$, proving that the allocation $X$ is also EEFX.

Assume otherwise. Then, by definition, there exists $j_1 \in [n] \setminus \{i\}$ such that $Y_{j_1} \neq \emptyset$ and
\begin{equation}\label{eq:not-efx}
v_i(X_i) < v_i(Y_{j_1})-v_i(g),
\end{equation}
where $g$ is the item in $Y_{j_1}$ for which agent $i$ has the minimum non-zero value.

Now, assume that $v_i(Y_j) > v_i(X_i)$ for every $j \in [n] \setminus \{i\}$.
Let $Z' = (Z'_1, \ldots, Z'_n)$ be the allocation obtained from $Y$ after removing item $g$ from bundle $Y_{j_1}$ and adding it to bundle $Y_i$.
Then $v_i(Z'_j) > v_i(X_i) \ge \mms_i$ for every $j \in [n]$.
(For $j \in [n] \setminus \{i,j_1\}$, this follows from our assumption. For $j=i$, this follows from $v_i(g) > 0$. For $j=j_1$, this follows from \cref{eq:not-efx} since $v_i(Z'_{j_1}) = v_i(Y_{j_1}) - v_i(g) > v_i(X_i)$.)
The existence of allocation $Z'$ contradicts the fact that allocation $X$ is MMS.

So, there must be $j_2 \in [n] \setminus \{i,j_1\}$ so that $v_i(Y_{j_2}) \le v_i(X_i)$.
Now, consider the allocation $Z'' = (Z''_1, \ldots, Z''_n)$ obtained after removing item $g$ from bundle $Y_{j_1}$ and adding it to bundle $Y_{j_2}$.
Notice that $v_i(Z''_j) = v_i(Y_j)$ for $j \in [n] \setminus \{j_1,j_2\}$,
$v_i(Z''_{j_1}) = v_i(Y_{j_1}) - v_i(g) > v_i(X_i) \ge v_i(Y_{j_2})$,
and $v_i(Z''_{j_2}) = v_i(Y_{j_2}) + v_i(g) > v_i(Y_{j_2})$.
Hence, $s^{Z'',i}$ is lexicographically larger than $s^{Y,i}$ and, furthermore, $Z''_i=X_i$, contradicting the assumption on $Y$. This completes the proof that $Y$ is agent $i$'s EEFX certificate for bundle $X_i$.
\end{proof}

\Cref{thm:mms_eefx} can be extended to the setting where agents' valuation functions may be non-additive.
See Appendix~\ref{sec:efx-defn} for details.
The implication of \cref{thm:mms_eefx} is strict. As we show in the next section, \eefx\ allocations always exist. This is not the case for \mms, as \citet{kurokawa2018fair} have proved. Then, any \eefx\ allocation in their counter-example instance cannot be \mms.

We conclude the section by establishing a connection between \mxs\ and \propOne.

\begin{theorem}[$\mxs \Rightarrow \propOne$]
\label{thm:mxs_prop1}
An \mxs\ allocation in a fair division instance with additive valuations is also \propOne.
\end{theorem}
\begin{proof}
Consider a fair division instance and let $X=(X_1, \ldots, X_n)$ be an \mxs\ allocation. Clearly, the proportionality and, consequently, the \propOne\ constraints are satisfied for every agent $i$ with $\mxs_i=\ps_i$. Now, consider an agent $i \in [n]$ with $\mxs_i<\ps_i$; we will show that $X$ satisfies the \propOne\ constraints for agent $i$ as well.

Assume otherwise that
\begin{equation}\label{eq:ps}
v_i(X_i) < \ps_i - v_i(g) \, ,
\end{equation}
for every item $g \in [m] \setminus X_i$. Let $Y=(Y_1, \ldots, Y_n)$ be an allocation in $\sefx_i$ such that $v_i(Y_i)=\mxs_i<\ps_i$. Since $\sum_{j \in [n]}v_i(Y_j)= n \cdot \ps_i$, there exists $k \in [n] \setminus \{i\}$ such that $v_i(Y_k) >\ps_i$. By \cref{eq:ps}, we have $v_i(Y_k)>v_i(X_i)$, meaning that there exists an item $g^*$ that belongs to $Y_k$ but not to $X_i$. By the definition of $\sefx_i$, we have
$\mxs_i = v_i(Y_i) \geq v_i(Y_k)-v_i(g^*) > \ps_i-v_i(g^*)$
and, using \cref{eq:ps}, we get
\[ v_i(X_i) < \ps_i-v_i(g^*) < \mxs_i, \]
contradicting the fact that allocation $X$ is \mxs.
\end{proof}

Again, the opposite implication is not true.

\begin{example}[$\propOne \nRightarrow \mxs$]
Consider the fair division instance with five goods and two agents with identical additive valuations.
The first two goods have value 4 each, and the last three goods have value 1 each.
It is easy to see that each agent's minimum EFX share is 5 and proportional share is $11/2$.
The allocation $(\{1\}, \{2, 3, 4, 5\})$ is PROP1 but not MXS.
\end{example}

\section{Existence and Efficient Computation of EEFX Allocations}
\label{sec:eefx}

We now present our main result.

\begin{theorem}\label{thm:eefx}
In any fair division instance where agents have additive valuations,
an \eefx\ allocation exists and can be computed in polynomial time.
\end{theorem}

We will prove \cref{thm:eefx} using \cref{alg:eefx}. We remark that this is essentially identical to the algorithm proposed by \citet{barman2020approximation} to compute $2/3$-\mms\ allocations. However, proving that the algorithm returns \eefx\ allocations requires additional arguments. Interestingly, our analysis is relatively simple.

\begin{algorithm}[t!]
\caption{Computing \eefx\ allocations}
\label{alg:eefx}
\begin{algorithmic}[1]
\Require A fair division instance $\instance$.
\Ensure An allocation $X$ in $\instance$.
\State $\ordInst \gets \ord(\instance)$
\State $\Xord \gets \celim(\ordInst)$
\State $L \gets \psequence(\ordInst, \Xord)$
\State $X \gets \pick(\instance, L)$
\State \Return $X$
\end{algorithmic}
\end{algorithm}

\Cref{alg:eefx} takes as input a fair division instance $\instance$ consisting of $m$ items and $n$ agents with valuations $\{v_i\}_{i\in [n]}$ and works as follows:
\begin{itemize}
\item It first (Step 1) executes routine $\ord$, which creates an ordered instance $\ordInst$ from instance $\instance$.
    Each agent $i$'s valuation function in $\ordInst$ is $\vord_i$ (c.f.~\cref{sec:prelims:order}).
\item Next, in Step 2, the algorithm executes the envy cycle elimination algorithm, for ordered instances, of \citet{lipton2004approximately} on instance $\ordInst$ to get an intermediate allocation $\Xord=(\Xord_1, \ldots, \Xord_n)$.
    The items are allocated in the order $1, 2, \ldots, m$.
\item Then, in Step 3, the routine $\psequence$ takes as input instance $\ordInst$ and allocation $\Xord$ and computes the picking sequence $L = [i_1, \ldots, i_m]$ as follows. For $j \in [m]$, $i_j$ is the agent who gets item $j$ in allocation $\Xord$, i.e., $j \in \Xord_{i_j}$.
\item Finally, in Step 4, the routine $\pick$ is executed with input the instance $\instance$ and the picking sequence $L$ to compute the allocation $X = (X_1, \ldots, X_n)$ as follows. $\pick$ runs $m$ rounds, one for each item. In round $j$, agent $i_j$ picks the smallest-ranked item that has not been allocated to any agent in rounds $1, 2, \ldots, j-1$.
\end{itemize}
\Cref{alg:eefx} returns allocation $X$ as output.

\Cref{alg:eefx} clearly runs in polynomial time. To complete the proof of \cref{thm:eefx}, we will show (in \cref{lem:eefx}) that allocation $X$ is \eefx. To do so, we will exploit two crucial properties maintained by the algorithm. The first one follows by a result of
\citet{plaut2020almost} and \citet{barman2020approximation}, who proved that the application of the envy cycle elimination algorithm on ordered fair division instances produces an \efx\ allocation.
The standard Envy Cycle Elimination algorithm starts with an empty allocation $X$. At each round, one available good $g$ is allocated to an agent $i$, i.e., $X_i \gets X_i \cup \set{g}$, if $i$ is not envied by any other agent. To ensure the existence of such an agent, at the beginning of each round, the envy graph corresponding to the partial allocation $X$ is built. In the envy graph, the nodes are the agents and there is a directed edge from $i$ to $j$ if $i$ envies $j$ in $X$. Source nodes are agents not envied by anyone, if a source does not exist it suffices to update $X$ by trading bundles along one of the existing cycles, that is, for each directed edge ${i,j}$ in the selected cycle, agent $i$ receives the bundle $X_j$. This process is repeated until a source node exists. The Envy Cycle Elimination algorithm guarantees $\ef1$ even for monotone valuations. For the specific case of ordered valuation, if during the $k\Th$ round the allocated good is the $k\Th$ most preferred, according to the common ranking of the agents, the outcome is also $\efx$. See Appendix~\ref{sec:cancelable-extra:ece} for further details.

\begin{lemma}[\citet{plaut2020almost}, \cite{barman2020approximation}]\label{lem:cycle-elim-of-ordered-instances-yields-efx}
The allocation $\Xord$ of instance $\ordInst$ is \efx.
\end{lemma}

The second crucial property is given by the next lemma.
Intuitively, it says that for each agent $i$, when we switch from $\Xord$ to $X$,
every item in her own bundle gets \emph{better},
and every item outside her bundle gets \emph{worse}.
The terms \emph{better} and \emph{worse} are formally defined in terms of
the rank function $r_i$ (c.f.~\cref{sec:prelims:order}).

\begin{lemma}\label{lem:properties-of-bijection}
For every agent $i\in [n]$, there exists a bijection $\pi_i: [m] \to [m]$ such that
\begin{itemize}
\item for every $j \in \Xord_i$, we have $\pi_i(j) \in X_i$ and $r_i(\pi_i(j)) \le j$.
\item for every $j \in [m] \setminus \Xord_i$, we have $\pi_i(j) \in [m] \setminus X_i$ and $r_i(\pi_i(j)) \ge j$.
\end{itemize}
\end{lemma}
\begin{proof}
Fix an agent $i$. Let $m_i \defeq |\Xord_i|$. Then $|X_i| = m_i$,
since agent $i$ picks an item in each round $j \in \Xord_i$ of $\pick$.

For every $k \in [m_i]$, let $\gord_k$ be agent $i$'s $k\Th$-favorite item in $\Xord_i$,
and let $g_k$ be agent $i$'s $k\Th$ favorite item in $X_i$.
Hence, $\gord_1 < \ldots < \gord_{m_i}$, and $r_i(g_1) < \ldots < r_i(g_{m_i})$.
Define $\pi_i(\gord_k) = g_k$ for all $k \in [m_i]$.

For every $k \in [m - m_i]$, let $\hord_k$ be agent $i$'s $k\Th$-favorite item in $[m] \setminus \Xord_i$,
and let $h_k$ be agent $i$'s $k\Th$ favorite item in $[m] \setminus X_i$.
Hence, $\hord_1 < \ldots < \hord_{m-m_i}$, and $r_i(h_1) < \ldots < r_i(h_{m-m_i})$.
Define $\pi_i(\hord_k) = h_k$ for all $k \in [m-m_i]$.
Hence, $\pi_i$ is a bijection from $[m]$ to $[m]$.

For each $k \in [m_i]$, in round $\gord_k$ of $\pick$, agent $i$ picks $g_k$,
which is their favorite remaining item in that round.
Since $\gord_k - 1$ items have been picked before that, $g_k$ is among their
top $\gord_k$ favorite items in $[m]$. Hence, $r_i(g_k) \le \gord_k$.
Therefore, $r_i(\pi_i(j)) \le j$ for all $j \in \Xord_i$.

Suppose $r_i(h_k) < \hord_k$ for some $k \in [m-m_i]$. $[\hord_k]$ can be partitioned into sets
$\Gord \defeq [\hord_k] \cap \Xord_i = \{\gord_1, \gord_2, \ldots, \gord_{|\Gord|}\}$
and $\Hord \defeq [\hord_k] \setminus \Xord_i = \{\hord_1, \hord_2, \ldots, \hord_k\}$.
For any $p \le k$, we get $r_i(h_p) \le r_i(h_k) < \hord_k$.
For any $\gord_q \in \Gord$, we have $r_i(g_q) \le \gord_q < \hord_k$
(by part 1 of this lemma).
Hence, $r_i(\pi_i(j)) < \hord_k$ for all $j \in [\hord_k]$.
Hence, $r_i \cdot \pi_i$ is a bijection that maps $[\hord_k]$ to $[\hord_k-1]$, which is a contradiction.
Therefore, $r_i(h_k) \ge \hord_k$ for all $k \in [m-m_i]$.
Hence, $r_i(\pi_i(j)) \ge j$ for all $j \in [m] \setminus \Xord_i$.
\end{proof}

We are now ready to complete the proof of \cref{thm:eefx} by proving the next lemma.

\begin{lemma}\label{lem:eefx}
Allocation $X$ is \eefx{} when agents have additive valuations.
\end{lemma}
\begin{proof}
Consider agent $i\in [n]$ and let $\pi_i$ be the bijection defined in \cref{lem:properties-of-bijection}. Define the allocation $Y$ with $Y_j = \{\pi_i(g): g \in \Xord_j\}$ for $j \in [n]$. Since $\pi_i$ is a bijection, allocation $Y$ is well-defined. Also, by \cref{lem:properties-of-bijection}, $Y_i = X_i$. We will~prove that $Y$ is an \eefx\ certificate for agent $i$ with bundle $X_i$.

Let $j \in [n] \setminus \{i\}$ and $g^*$ be the item of bundle $\Xord_j$ such that $\pi_i(g^*)$ (which, by definition, belongs to bundle $Y_j$) is the largest-ranked item of non-zero value according to $v_i$. Thus, proving $v_i(Y_i) \geq v_i(Y_j \setminus \{\pi_i(g^*)\})$ is enough to complete the proof.

Since $g^* \not\in \Xord_i$ and $v_i(\pi_i(g^*)) > 0$, \cref{lem:properties-of-bijection} implies that $\pi_i(g^*) \not\in X_i$ and $\vord_i(g^*) \ge \vord_i(r_i(\pi_i(g^*))) = v_i(\pi_i(g^*)) > 0$. Then, the fact that $\Xord$ is \efx\ for instance $\ordInst$ (from \cref{lem:cycle-elim-of-ordered-instances-yields-efx}) implies
\begin{equation}\label{eq:X-prime-is-efx}
\vord_i(\Xord_i) \geq \vord_i(\Xord_j \setminus \{g^*\}).
\end{equation}
Now, the properties of $\pi_i$ from \cref{lem:properties-of-bijection} yield
\begin{equation}\label{eq:lower-bound-on-Y_i}
v_i(Y_i) = \sum_{g \in \Xord_i}{\vord_i(r_i(\pi_i(g)))} \geq \sum_{g\in \Xord_i}{\vord_i(g)} = \vord_i(\Xord_i)
\end{equation}
and
\ifx\VERSION\versionOpre
\begin{align}\nonumber
\vord_i(\Xord_j \setminus \{g^*\}) &= \sum_{g \in \Xord_j \setminus \{g^*\}}{\vord_i(g)}\\
\nonumber   &\geq \sum_{g \in \Xord_j \setminus \{g^*\}}{\vord_i(r_i(\pi_i(g)))}\\\label{eq:upper-bound-on-Y_j}
&= v_i(Y_j \setminus \{\pi_i(g^*)\}).
\end{align}
\else
\begin{equation}
\label{eq:upper-bound-on-Y_j}
\vord_i(\Xord_j \setminus \{g^*\})
    = \sum_{g \in \Xord_j \setminus \{g^*\}}{\vord_i(g)}
    \ge \sum_{g \in \Xord_j \setminus \{g^*\}}{\vord_i(r_i(\pi_i(g)))}
    = v_i(Y_j \setminus \{\pi_i(g^*)\}).
\end{equation}
\fi
By applying equations \eqref{eq:lower-bound-on-Y_i}, \eqref{eq:X-prime-is-efx}, and \eqref{eq:upper-bound-on-Y_j},
we get the desired inequality $v_i(Y_i) \geq v_i(Y_j\setminus \{\pi_i(g^*)\})$.
\end{proof}

We remark that the complexity of verifying if a given allocation is \eefx\ is currently an open problem. Fortunately, \cref{lem:eefx} implies that agents can \emph{trust} that the allocation computed by \cref{alg:eefx} is \eefx. Moreover, note that the proof of \cref{lem:eefx} shows that \cref{alg:eefx} can be used to explicitly provide to each agent her \eefx\ certificate in polynomial time.

By \cref{thm:eefx_mxs,thm:eefx}, we obtain the next corollary.
Pareto-optimality follows since any Pareto-improvement of an \mxs\ allocation is \mxs\ as well.

\begin{corollary}\label{cor:mxs-does-exist}
In any fair division instance, there exists a Pareto-optimal \mxs\ allocation. Furthermore, an \mxs\ allocation can be computed in polynomial time.
\end{corollary}

\section{Computing the Minimum EFX Share}
\label{sec:hardness}

We now present a hardness result for computing the minimum \efx\ share, which may come as a surprise given the positive result in Corollary~\ref{cor:mxs-does-exist}. Fortunately, Algorithm~\ref{alg:eefx} computes \mxs\ allocations without computing the minimum \efx\ share of any agent at any point of its execution.

\begin{theorem}\label{thm:MFXvalueNPhard}
Computing the minimum \efx\ share of the agents in a fair division instance is $\np$-hard.
\end{theorem}

\begin{proof}
We prove the theorem by developing a polynomial-time reduction from the $\np$-hard problem of \bpartition\ \cite{GJ79}.

\begin{table}[!ht]\centering
\renewcommand{\arraystretch}{1.2}
\begin{tabular}{lc}
\toprule
\textbf{\bpartition}\\ \midrule
\textbf{Input:} A set $N= \{x_1, \dots, x_{2t}\}$ of positive integer \\values such that $\sum_{h=1}^{2t}x_h = 2T$, for $t,T\in \naturals$\\
\textbf{Problem:} Does there exist a balanced partition of $N$,\\
i.e., an equipartition $(S,N\setminus S)$
 of the elements in $N$ \\(i.e., $|S|=t$) such that $\sum_{h\in S}x_h = \sum_{h\in N\setminus S}x_h$? \\ \bottomrule
\end{tabular}
\end{table}
Starting from an instance $\phi$ of \bpartition\ with a set $N$ of elements and parameters $t$ and $T$, we construct a fair division instance $\instance(\phi)$ as follows. For $h=1, ..., 2t$, there is an {\em element item} corresponding to the element $x_h\in N$; there is also an {\em extra item} $2t+1$. There are two agents having identical valuations denoted by $v$; the value both agents have is $4T-x_h$ for the element item $h\in [2t]$ and $2T$ for the extra item $2t+1$. Our reduction is clearly polynomial-time.

For every allocation $X=(X_1,X_2)$ in $\instance(\phi)$, define the {\em induced partition} $(S_X,N\setminus S_X)$ as follows. Let $i\in \{1,2\}$ be such that $2t+1\in X_i$. Then, $S_X=\{x_h\in N: h\in X_i\}$. The next two lemmas state structural properties of \efx\ allocations and their induced partitions.

\begin{lemma}\label{lem:efx-implies-equipartition}
Let $X=(X_1,X_2)$ be an \efx\ allocation of $\instance(\phi)$. Then, its induced partition $(S_X,N\setminus S_X)$ is an equipartition of $N$.
\end{lemma}

\begin{proof}
Let $t_1$ and $t_2$ be the number of element items that bundles $X_1$ and $X_2$ have. Let $i\in \{1,2\}$ be such that the extra item $2t+1$ belongs to bundle $X_i$. Consequently, $|S_X|=t_i$ and $|N\setminus S_X|=t_{3-i}$. Since $X$ is \efx, bundle $X_{3-i}$ cannot be empty; let $g$ be the least valued item in $X_{3-i}$.

By the definition of valuations, we have
\begin{equation}
\sum_{x_h\in S_X}{x_h}-\sum_{x_h\in N\setminus S_X}{x_h} = (4t_i-4t_{3-i})T
\label{eq:basic-equation}
\ \  -v(X_i\setminus\{2t+1\})+v(X_{3-i}).
\end{equation}
Since agent $3-i$ is \efx-satisfied and $\sum_{x_h\in S_X}{x_h}\leq 2T$, equation (\ref{eq:basic-equation}) yields

$2T\geq \sum_{x_h\in S_X}{x_h}-\sum_{x_h\in N\setminus S_X}{x_h}
\geq (4t_i-4t_{3-i})T$,
i.e., $t_i-t_{3-i}\leq 1/2$. Also, since agent $i$ is \efx-satisfied, the facts $\sum_{x_h\in N\setminus S_X}{x_h}\leq 2T$ and $v(g)<4T$, and equation (\ref{eq:basic-equation}) yield

$-2T \leq (4t_i-4t_{3-i})T-v(X_i)+v(X_{3-i}\setminus \{g\}) +v(2t+1)+v(g) < (4t_i-4t_{3-i}+6)T$,
i.e., $t_i-t_{3-i}>-2$. Since $t_i+t_{3-i}=2t$ the difference $t_i-t_{3-i}$ must be an even integer, and $t_i=t_{3-i}$ is the only case allowed by the inequalities $-2<t_i-t_{3-i}\leq 1/2$, implying that $S_X$ is an equipartition.
\end{proof}

\begin{lemma}\label{lem:structure}
Let $X=(X_1,X_2)$ be an \efx\ allocation of $\instance(\phi)$ and $i\in \{1,2\}$ is such that $2t+1\in X_i$. Then,
$v(X_i)\geq v(X_{3-i})\geq v(X_i\setminus \{2t+1\})$.
\end{lemma}

\begin{proof}
Notice that the rightmost inequality follows since agent $3-i$ is \efx-satisfied. Now, by the definition of the valuations, the fact that the induced partition $(S_X,N\setminus S_X)$ of allocation $X$ is actually an equipartition (by \cref{lem:efx-implies-equipartition}), and since $\sum_{h\in S_x}{x_h}\leq 2T$, we have
\begin{align*}
v(X_i) &=2T+\sum_{x_h\in S_X}{(4T-x_h)}\\
&=2T+v(X_{3-i})+\sum_{x_h\in N\setminus S_X}{x_h}-\sum_{x_h\in S_X}{x_h}\\
&\geq v(X_{3-i}),
\end{align*}
thus proving the leftmost inequality as well.
\end{proof}

We next establish a connection between any \efx\ allocation and its induced equipartition.
\begin{lemma}\label{lem:connection-between-efx-allocations-and-equipartitions}
Let $X=(X_1,X_2)$ be an \efx\ allocation of $\instance(\phi)$ that induces the partition $(S_X,N\setminus S_X)$ of $N$. Then, $\min\{v(X_1),v(X_2)\}+\min\left\{\sum_{x_h\in S_X}{x_h},\sum_{x_h\in N\setminus S_X}{x_h}\right\}=4tT$.
\end{lemma}

\begin{proof}
Let $i\in \{1,2\}$ be such that $2t+1\in X_i$. Since both bundles $X_i$ and $X_{3-i}$ have $t$ element items each (by \cref{lem:efx-implies-equipartition}), the inequality $v(X_i)\geq v(X_{3-i})$ from \cref{lem:structure} implies that
\begin{align}\nonumber
\min\{v(X_1),v(X_2)\} &= v(X_{3-i})=\sum_{x_h\in N\setminus S_X}{(4T-x_h)}\\\label{eq:min-between-values}
&=4tT -\sum_{x_h\in N\setminus S_X}{x_h},
\end{align}
and the inequality $v(X_{3-i})\geq v(X_i\setminus \{2t+1\})$ from \cref{lem:structure} yields $\sum_{x_h\in N\setminus S_X}{(4T-x_h)}\geq \sum_{h\in S_X}{(4T-x_h)}$ and, equivalently,
\begin{align}\label{eq:min-between-partition-sides}
\min\left\{\sum_{x_h\in S_X}{x_h},\sum_{x_h\in N\setminus S_X}{x_h}\right\}=\sum_{x_h\in N\setminus S_X}{x_h}.
\end{align}
The lemma follows by equations (\ref{eq:min-between-values}) and (\ref{eq:min-between-partition-sides}).
\end{proof}

\Cref{lem:connection-between-efx-allocations-and-equipartitions} implies that there exists an \efx\ allocation $X$ with $\min\{v(X_1),v(X_2)\}=(4t-1)T$ (and, thus, the minimum \efx\ share of both agents is at most $(4t-1)T$) if and only if its induced partition is balanced. To complete the proof of correctness for our reduction, we need to prove that every balanced partition is the induced partition of some \efx\ allocation; we do so in the following.

Consider the balanced partition $(S,N\setminus S)$, i.e., $\sum_{x_h\in S}{x_h}= \sum_{x_h\in N\setminus S_X}{x_h}$. Let us consider the allocation $X=(X_1,X_2)$ where $X_1 = \{h\in [2t]: x_h\in S\}\cup \{2t+1\}$ and $X_2=\{h\in [2t]: x_h\in N\setminus S\}$. Note that it is straightforward to verify that the allocation $X$ has the equipartition $S$ as induced partition. To complete the proof, we will show that $X$ is \efx.

For the sake of contradiction, assume otherwise that $X$ is not \efx. Since $S$ is an equipartition, it is trivial to see that $v(X_1)\geq v(X_2)$. Therefore, the only possibility is that agent~$2$ is not \efx-satisfied. Then, the item $2t+1$ is the least-valued item in bundle $X_1$ and the fact that agent $2$ is not \efx-satisfied yields
\begin{align*}
0 &> v(X_2)-v(X_1\setminus \{2t+1\})\\
&=\sum_{x_h\in N\setminus S_X}{(4T-x_h)}-\sum_{x_h\in S}{(4T-x_h)}\\
&=4(t_2-t_1)T+\sum_{x_h\in N\setminus S_X}{x_h}-\sum_{x_h\in S}{x_h}.
\end{align*}
Thus, either $t_1\not=t_2$ and $(S,N\setminus S)$ is not an equipartition or $t_1=t_2$ but $\sum_{x_h\in N\setminus S_X}{x_h}<\sum_{x_h\in S}{x_h}$, meaning that $(S,N\setminus S)$ is an equipartition but not balanced. In any case, we obtain the desired contradiction.
\end{proof}

\section{Chores}
\label{sec:chores}

In this section, we study the problem of fairly dividing a set of chores, instead of goods. That is, we understand the computation of $\eefx$ allocations for negatively-valued chores and show that our results for the case of goods can be extended to this setting. We begin by defining the fairness notions for chores.

\begin{definition}[Envy-freeness up to any item (EFX)]
\label{defn:efx-chores}
An agent $i$ is \emph{EFX-satisfied} by an allocation $X = (X_1, \ldots, X_n)$, if for any other agent $k$,
and any chore $c \in X_i$ such that $v_i(c) > 0$, it holds that $v_i(X_i \setminus \set{c}) \ge v_i(X_k)$.
The allocation $X$ is \emph{EFX} if every agent is EFX-satisfied by $X$.
\end{definition}

We can now similarly define the concept of $\eefx$ allocations for chores.
\begin{definition}[Epistemic \efx\ (\eefx) and \eefx\ certificates]
For a fair division instance with chores, an allocation $X = (X_1, X_2, \ldots, X_n)$ is called {\em epistemic \efx\ (\eefx)}
if for every agent $i \in \agents$, there exists an allocation $Y = (Y_1, Y_2, \ldots, Y_n)$ such that
$Y_i = X_i$ and agent $i$ is \efx-satisfied with $Y$.
We call such an allocation $Y$ an \emph{\eefx\ certificate} of agent $i$ for bundle $X_i$.
\end{definition}

We start by noticing some key distinctions in the case of chores. Here, the item to be virtually removed so as to eliminate the envy is in the bundle of the envious, and not of the envied, agent. Although this may seem like a minor change, this difference is crucial and leads to distinct results in the goods and chores settings.
As an example, when running the envy cycle elimination we must assign items to some envied agent, that is, a sink node, rather than a source node. Nonetheless, the envy cycle elimination algorithm does not necessarily guarantee $\ef1$ allocation in the case of chores (\citet{bhaskar2021}). Here, when trading bundles along an envy-cycle (to resolve it), each participating agent in the cycle receives a strictly better bundle; however, if she remains envious of some other agent, it might not be true that by removing a chore from her bundle, the envy would be eliminated. Therefore, for ordered instances with chores, \Cref{lem:cycle-elim-of-ordered-instances-yields-efx} does not necessarily hold true.
Fortunately, it has been proven that a variant of the envy cycle elimination algorithm, the so-called top-trading envy cycle elimination ($\ttcelim$) will find an $\ef1$ allocation for chores; see~\citet{bhaskar2021} for further details. Therefore, for finding \efx\ allocations for ordered instances with chores, we can use the $\ttcelim$ rather than $\celim$ where we rank the chores from the least to the most preferred.
With the scope of obtaining \eefx\ in the context of chores, we replace $\celim$ with $\ttcelim$ in Algorithm~\ref{alg:eefx}. We call this as the \emph{modified} Algorithm~\ref{alg:eefx}. This change guarantees that \Cref{lem:cycle-elim-of-ordered-instances-yields-efx} holds true in the case of chores.
Moreover, in \citet{barman2020approximation}, where \efx\ is out of consideration, the authors have shown that any run of the envy cycle elimination on ordered instances with chores guarantees a $4/3$-\mms\ allocation; this shows that our algorithm guarantees a 4/3-\mms\ allocation as well.

\begin{lemma} \label{lem:chores_ordered}
There exists a polynomial-time algorithm to compute $\efx$ allocations of chores for ordered instances with additive valuations.
\end{lemma}

Next, it is easy to verify that \Cref{lem:properties-of-bijection} holds true in the case of chores. Let us therefore prove the main theorem where we show the existence and tractability of $\eefx$ allocations of chores among agents with additive valuations. Recall that we rank items (chores) from the least to the most preferred to create the ordered instance $\ordInst$.

\begin{lemma}\label{lem:eefx-chores}
For fair division instances with chores, the output allocation $X$ (of modified Algorithm~\ref{alg:eefx}) is \eefx\ when agents have additive valuations.
\end{lemma}

\begin{proof}
Consider an agent $i\in [n]$ and let $\pi_i$ be the bijection defined in \cref{lem:properties-of-bijection}. Define the allocation $Y$ with $Y_j = \{\pi_i(c): c \in \Xord_j\}$ for $j \in [n]$. Since $\pi_i$ is a bijection, allocation $Y$ is well-defined. Also, by \cref{lem:properties-of-bijection}, $Y_i = X_i$. We will~prove that $Y$ is an \eefx\ certificate for agent $i$ with bundle $X_i$.

Let $c^*$ be the item of bundle $\Xord_i$ such that $\pi_i(c^*)$ is the lowest-ranked chore (i.e., the most preferred chore in bundle $Y_i=X_i$) according to $v_i$. Thus, proving $v_i(Y_i \setminus \{\pi_i(c^*)\}) \geq v_i(Y_j)$ is enough to complete the proof.
Since $c^*\in \Xord_i$ and $v_i(\pi_i(c^*)) < 0$, \cref{lem:properties-of-bijection} implies that $\pi_i(c^*) \in X_i$ and $v_i(\pi_i(c^*)) < 0$. Then, the fact that $\Xord$ is \efx\ for instance $\ordInst$ (from \cref{lem:cycle-elim-of-ordered-instances-yields-efx}) implies
\begin{equation}\label{eq:X-prime-is-efx-chore}
v_i(\Xord_i\setminus \{c^*\}) \geq v_i(\Xord_j).
\end{equation}
Now, the properties of $\pi_i$ from \cref{lem:properties-of-bijection} yield
\begin{align}\label{eq:lower-bound-on-Y_i-chore}
v_i(Y_i \setminus \{\pi_i(c^*)\}) &= \sum_{c \in \Xord_i \setminus \{c^*\}}{v_i(\pi_i(c)))} \nonumber
\\ &\geq \sum_{c\in \Xord_i \setminus \{c^*\}}{v_i(c)} = v_i(\Xord_i \setminus \{c^*\})
\end{align}
and
\begin{equation}\label{eq:upper-bound-on-Y_j-chore}
v_i(\Xord_j) = \sum_{c \in \Xord_j}{v_i(c)} \geq \sum_{c \in \Xord_j}{v_i(r_i(\pi_i(c)))} = v_i(Y_j)
\end{equation}

By applying equations (\ref{eq:lower-bound-on-Y_i-chore}), (\ref{eq:X-prime-is-efx-chore}), and (\ref{eq:upper-bound-on-Y_j-chore}), we get the desired inequality $v_i(Y_i \setminus \{\pi_i(c^*)\} ) \geq v_i(Y_j)$. Hence, the claim stands proven.
\end{proof}

We remark that the complexity of verifying if a given allocation is \eefx\ is an $\np$-complete problem (as we prove next in Theorem~\ref{thm:eefx-hard-chores}). Fortunately, \cref{lem:eefx-chores} implies that agents can \emph{trust} that the allocation computed by \cref{alg:eefx} is \eefx. Moreover, note that the proof of \cref{lem:eefx-chores} shows that \cref{alg:eefx} can be used to explicitly provide to each agent her \eefx\ certificate in polynomial time. We therefore obtain the following result.

\begin{theorem}\label{thm:eefx-chores}
In any fair division instance with chores where agents have additive valuations, an \eefx\ allocation exists and can be computed in polynomial time.
\end{theorem}

Finally, we prove that, given an arbitrary allocation of chores among three agents with additive valuations, it is $\np$-complete to decide if it is $\eefx$.

\begin{theorem} \label{thm:eefx-hard-chores}
It is \np-complete to determine if an allocation $X$ is \eefx\ for a given agent, even if there are only three agents with additive valuations.
\end{theorem}
\begin{proof}
We prove the stated theorem by developing a polynomial time reduction from the $\np$-complete problem of \partition, defined as below. Starting with an instance $\phi$ of $\partition$, we will construct a fair-division instance $\instance(\phi)$ with three agents such that there exists a partition in $\phi$ if and only if the corresponding allocation in $\instance(\phi)$ is $\eefx$ for the first agent.

\begin{table}[ht]\centering
\renewcommand{\arraystretch}{1.4}
\begin{tabular}{lc}
\toprule
 \partition\\ \midrule
Given: A set $N=\set{x_1, \dots, x_{m}}$ of positive integers such that $\sum_{h=1}^{n}x_h = 2T$ \\
Question: Is there a partition of $N$, $(S, N\setminus S)$, such that $ \sum_{x_h\in S}x_h = \sum_{h\in N\setminus S}x_h$? \\ \bottomrule
\end{tabular}
\end{table}

The instance c will consist of three agents and $m+2$ chores. We define the valuation $v_1$ of agent $1$ and consider some arbitrary valuations for the remaining two agents. We set $v_1(h)=-x_h$ for $h \in [m]$, and $v_1(m+1)=v_1(m+2)=-T$. Now, consider the allocation $X$ where $X_1=\set{m+1,m+2}$ having a vale of $-2T$ for agent $1$. We will show that $X$ is \eefx\ for agent $1$ in $\instance(\phi)$ if and only if there exists a partition in $\phi$.

Note that $X$ is \eefx\ for agent $1$ if and only if there exists a reallocation $Y = \{Y_1,Y_2,Y_3\}$ where $Y_1=X_1$ and agent $1$ is \efx\ in $Y$. That is, we have $v_1(Y_1\setminus \set{c})\geq \max\set{v_1(Y_2),v_1(Y_3)}$ for each $c\in\set{m+1,m+2}$. Since, for any $c\in Y_1$, $v_1(Y_1\setminus \set{c})=-T$ and $\max\set{v_1(Y_2),v_1(Y_3)}\geq - T$, a reallocation $Y$ exists if and only if $\max\set{v_1(Y_2),v_1(Y_3)}= -T$. This in turn is possible if and only if $v_1(Y_2)=v_1(Y_3)$, i.e., if and only if there exists a partition in $\phi$. This is so as there is a one to one correspondence between any partition of $[m]$ into two bundles $Y_2, Y_3$ in $\instance(\phi)$ and a partition $(S, N\setminus S)$ for $\phi$.
\end{proof}

\subsection{Relations to Other Fairness Concepts}
\label{sec:chores:relation}

We study connections between \eefx\ and \mxs\ with previously well-studied notions of fairness in the literature. We show that these connections differ significantly between goods and chores.
The implications $\mms \Rightarrow\eefx$ and $\mxs \Rightarrow \propOne$ hold for goods but not for chores.
On the other hand, we get $\eefx \Rightarrow \propx$ for chores.

We begin with a simple result showing how \eefx\ and \mxs\ allocations are related.

\begin{theorem}[$\eefx \Rightarrow \mxs$]
\label{thm:eefx-mxs-chores}
An \eefx\ allocation of chores is also \mxs.
\end{theorem}
\begin{proof}
Let $X = (X_1, \ldots, X_n)$ denote an $\eefx$ allocation in a fair division instance.
Fixing an agent $i \in [n]$, we will prove that $v_i(X_i)$ is at least as high as her minimum \efx\ share.
By definition, since $X$ is an \eefx\ allocation, there must exist an \eefx\ certificate $Y = (Y_1, \ldots, Y_n)$ for agent $i$ such that $Y_i = X_i$ and $Y \in \sefx_i$.
Therefore, we can write
\[ v_i(Y_i) \ge \min\limits_{Z \in \sefx_i}v_i(Z_i) = \mxs_i. \]
Since, the bundles $X_i$ and $Y_i$ are identical, we obtain $v_i(X_i) \geq \mxs_i$.
This holds for every agent $i \in [n]$, so $X$ is an \mxs\ allocation.
\end{proof}

By \cref{thm:eefx-mxs-chores,thm:eefx-chores}, we obtain the next corollary.
Pareto-optimality follows since any Pareto-improvement of an \mxs\ allocation is \mxs\ as well.

\begin{corollary}\label{cor:mxs-does-exist-chores}
In any fair division instance, there exists a Pareto-optimal \mxs\ allocation.
Furthermore, an \mxs\ allocation can be computed in polynomial time.
\end{corollary}

The implication in \cref{thm:eefx-mxs-chores} is strict
(even for 2 agents having identical valuations),
as the following example shows.

\begin{example}[$\mxs \nRightarrow \eefx$]
\label{ex:eefx-nimpl-mxs-chores}
Consider a fair division instance with 2 agents having identical additive valuations over 7 chores.
The first 2 chores have a disutility of $4$ each, and the remaining chores have a disutility of 1 each.

Agent 1 is EFX-satisfied by the allocation $([2], [7] \setminus [2])$, so $\mxs_1 \le -8$.
Agent 2 is EFX-satisfied by the allocation $([7] \setminus [2], [2])$, so $\mxs_2 \le -8$.
The allocation $X \defeq (\{1, 7\}, \{2, 3, 4, 5, 6\})$ is MXS, since each agent gets a bundle of value at least $-8$.
However, agent 2 not EFX-satisfied by $X$, since her disutility for her own bundle is 7 after removing good 6,
but her disutility for agent 1's bundle is 5.
Hence, $X$ is not an EEFX allocation.
\end{example}

We showed in \cref{thm:mms_eefx} that an MMS allocation of goods is also an \eefx\ allocation.
Surprisingly, this is not true for chores.

\begin{example}[$\mms \nRightarrow \eefx$]
\label{ex:mms-nimpl-efx-chores}
Consider a fair division instance with 3 agents and 5 chores.
The agents have identical valuations, and the chores have disutilities $40$, $30$, $30$, $30$, $10$.

Each agent's maximin share is $60$. Let $X \defeq (\{1, 5\}, \{2, 3\}, \{4\})$, i.e.,
agent 1 gets the chores of disutilities $40$ and $10$,
agent 2 gets two chores of disutilities $30$ each,
and agent 3 gets a single chore of disutility $30$.
Then $X$ is an MMS allocation. However, $X$ is not an EEFX allocation
because in any EEFX-certificate for agent 1,
some agent gets a bundle of value $30$.
\end{example}

We now define the notions \propx\ and \propOne,
and show their connections to \eefx\ and \mxs, respectively.

\begin{definition}[\propx\ and \propOne]
\label{defn:prop1-chores}
For a fair division instance, we define agent $i$'s \emph{proportionality threshold} as
$\ps_i := v_i([m]) / n$.
An allocation $X = (X_1, \ldots, X_n)$ is \emph{proportional up to any item} ($\propx$) if
$\max_{c \in X_i: v_i(c) < 0} |v_i(X_i \setminus \{c\})| \le |\ps_i|$ for every agent $i \in [n]$.
An allocation $X = (X_1, \ldots, X_n)$ is \emph{proportional up to one item} ($\propOne$) if
$\min_{c \in X_i} |v_i(X_i \setminus \{c\})| \le |\ps_i|$ for every agent $i \in [n]$.
\end{definition}

It is easy to see that a \propx\ allocation of chores is also a \propOne\ allocation
when valuations are additive.

\begin{theorem}[$\eefx \Rightarrow \propx$]
\label{thm:eefx-impl-propx-chores}
An \eefx\ allocation $X$ of chores is also a \propx\ allocation.
\end{theorem}
\begin{proof}
Our proof is very similar to Lemma 2.1 of \cite{li2022almost},
where they show that an \efx{} allocation of chores is also \propx{}.

Pick an arbitrary agent $i$. If $v_i(X_i) = 0$, then $|v_i(X_i)| \le |\ps_i|$.
Now assume $v_i(X_i) < 0$. Since valuations are additive, this means that
for some chore $c \in X_i$, we have $v_i(c) < 0$.
Let $Y$ be agent $i$'s EEFX certificate for $X$.
Let $\chat \in \argmin_{c \in Y_i: v_i(c) < 0} |v_i(c)|$.
Then for any agent $j \neq i$, we have $|v_i(Y_i \setminus \{\chat\})| \le |v_i(Y_j)|$.
Also, $|v_i(Y_i \setminus \{\chat\})| < |v_i(Y_i)|$.
Hence, on taking the mean of these $n$ inequalities, we get
\[ |v_i(Y_i \setminus \{\chat\})| < \frac{1}{n}\sum_{j=1}^n |v_i(Y_j)| = |\ps_i|. \]
Since $X_i = Y_i$, we get
\[ \max_{c \in X_i: v_i(c) < 0} |v_i(X_i \setminus \{c\})|
= v_i(Y_i \setminus \{\chat\}) < |\ps_i|. \]
Hence, $X$ is \propx.
\end{proof}

\begin{example}[$\mxs \nRightarrow \propOne$]
\label{ex:mxs-nimpl-prop1-chores}
Consider a fair division instance with 3 agents having identical additive disutilities.
Let there be 2 chores of disutility $4$ each (big chores) and 10 chores of disutility $1$ each (small chores).
Let $X$ be an allocation where agent 1 has 8 small chores,
agent 2 has a big chore and a small chore, and agent 3 has a big chore and a small chore.
The proportional share is $-6$, so agent 1 is not \propOne-satisfied by $X$.
We will show that $X$ is an \mxs\ allocation.

Let $Y$ be an allocation where agent 1 has both big chores,
and agents 2 and 3 have 5 small chores each.
Then the disutility of $X_i$ and $Y_i$ is the same for every agent $i$,
and each agent is \efx-satisifed in $Y$.
Hence, $X$ is an \mxs\ allocation.
\end{example}

\section{Cancelable Valuations}
\label{sec:cancelable}

Although we primarily focus on additive valuations in this work, we show, in this section, that EEFX allocations exist
and can be efficiently computed, even when agents have \emph{cancelable} valuations.
This class was introduced by \citet{berger2021almost}, and strictly generalizes the class of additive valuations.
Cancelable valuations also generalize unit demand valuations ($v_i(S) \defeq \max\{g \in S: v_i(g)\}$),
budget additive valuations ($v_i(S) \defeq \max\{B_i, \sum_{g \in S} v_i(g)\}$ for some $B_i$),
and multiplicative valuations ($v_i(S) \defeq \prod_{g \in S} v_i(g)$).

\begin{definition}
\label{defn:cancelable}
A function $u: 2^{[m]} \to \mathbb{R}$ is called \emph{cancelable} if
for all $g \in [m]$ and $S_1, S_2 \subseteq [m] \setminus \{g\}$,
we have $u(S_1 \cup \{g\}) > u(S_2 \cup \{g\}) \Rightarrow u(S_1) > u(S_2)$.
\end{definition}

We first state an equivalent definition of cancelable valuations.
Roughly, it says that if the valuation function is cancelable, then among any two sets,
the set having a larger value also has a greater or equal \emph{marginal value} with respect to any set $T$. We refer the readers to Appendix~\ref{sec:cancelable-extra:defn-equiv} for the proof of Lemma~\ref{thm:canc-defn-equiv}.

\begin{restatable}{lemma}{thmCancDefnEquiv}
\label{thm:canc-defn-equiv}
Let $u: 2^{[m]} \to \mathbb{R}$ be a function. Then $u$ is cancelable iff
for all $T \subseteq [m]$ and $S_1, S_2 \subseteq [m] \setminus T$,
we have $u(S_1) \ge u(S_2) \Rightarrow u(S_1 \cup T) \ge u(S_2 \cup T)$.
\end{restatable}

Before computing EEFX allocations when agents have cancelable valuations,
we first need to extend the definitions of EFX and EEFX to non-additive valuations.
This is tricky (see Appendix~\ref{sec:efx-defn} for details), so instead,
we use the slightly stronger fairness notions \efxZero\ and \eefxZero.

\begin{restatable}[\efxZero]{definition}{defnEfxZero}
An agent $i$ is \efxZero\emph{-satisfied} by allocation $X = (X_1, \ldots, X_n)$
if for every other agent $j$, we have
$\displaystyle v_i(X_i) \ge \max_{g \in X_j} v_i(X_j \setminus \{g\})$.
The allocation $X$ is \efxZero\ if every agent is \efxZero-satisfied by $X$.
\end{restatable}

We can define \eefxZero\ using \efxZero\ just like how we defined \eefx\ using \efx.

Next, we show two important properties of cancelable functions (in Lemmas~\ref{thm:cancelable-agg} and \ref{thm:cancelable-last})
that help us analyze our algorithm for computing \eefxZero\ allocations.

\subsection{Comparing Bundles using Ordered Cancelable Valuations}
\label{sec:cancelable:ordered}

A function $u: 2^{[m]} \to \mathbb{R}$ is called \emph{ordered} if
$u(\{1\}) \ge u(\{2\}) \ge \ldots \ge u(\{m\})$.
In \cref{thm:cancelable-agg}, we show that that for any two sets $S$ and $T$,
if each item in $S$ is \emph{better} than the corresponding item of $T$,
then $S$ is \emph{better} than $T$.
In \cref{thm:cancelable-last}, we show that for any set,
the strict subset of maximum value is the one where we remove the least-valued element.
Note that these results are trivial to prove for the special case where $u$ is additive.

\begin{lemma}
\label{thm:cancelable-agg}
Let $S, T \subseteq [m]$ and $\sigma: T \to S$ be a bijection where $\sigma(t) \le t$ for all $t \in T$. For an ordered cancelable function $u: 2^{[m]} \to \mathbb{R}$, all of the following statements hold true:

1. $u(S) \ge u(T)$.
\quad 2. $\displaystyle \max_{s \in S} u(S \setminus \{s\}) \ge \max_{t \in T} u(T \setminus \{t\})$.
\quad 3. $\displaystyle \min_{s \in S} u(S \setminus \{s\}) \ge \min_{t \in T} u(T \setminus \{t\})$.
\end{lemma}
\begin{noQedProof}
Since $\sigma: T \to S$ is a bijection, we have $|S| = |T|$.
Let $S \defeq \{s_1, \ldots, s_k\}$ and $T \defeq \{t_1, \ldots, t_k\}$
such that $\sigma(t_i) = s_i$. Then $s_i \le t_i$ for all $i \in [k]$.

For $i \in [k]$, define $Q_i \defeq \{s_1, \ldots, s_{i-1}\} \cup \{t_{i+1}, \ldots, t_k\}$.
We can rewrite $u(S) - u(T)$ as
\[ \sum_{i=1}^k (u(\{s_i\} \cup Q_i) - u(\{t_i\} \cup Q_i)). \]
Since $s_i \le t_i$ for all $i \in [k]$, and $u(1) \ge u(2) \ge \ldots \ge u(m)$,
we get $u(s_i) \ge u(t_i)$ for all $i \in [k]$.
Since $u$ is cancelable, we get $u(\{s_i\} \cup Q_i) \ge u(\{t_i\} \cup Q_i)$ for all $i \in [k]$.
Hence, $u(S) \ge u(T)$.

For any $T' \subseteq T$, define $\sigma(T') \defeq \{t \in T': \sigma(t)\}$.
Then by part 1 of the lemma, we get
\[ \max_{s \in S} u(S \setminus \{s\})
    = \max_{t \in T} u(\sigma(T \setminus \{t\}))
    \ge \max_{t \in T} u(T \setminus \{t\}), \textrm{ and} \]
\[ \min_{s \in S} u(S \setminus \{s\})
    = \min_{t \in T} u(\sigma(T \setminus \{t\}))
    \ge \min_{t \in T} u(T \setminus \{t\}).
    \tag*{\Halmos} \]
\end{noQedProof}

\begin{lemma}
\label{thm:cancelable-last}
Let $u: 2^{[m]} \to \mathbb{R}$ be an ordered cancelable function.
Let $S \subseteq [m]$ and $s^* \defeq \max(S)$ (i.e., $s$ is the largest-indexed element in $S$).
Then $u(S \setminus \{s^*\}) = \max_{s \in S} u(S \setminus \{s\})$.
\end{lemma}
\begin{proof}
Let $\shat \in \argmax_{s \in S} u(S \setminus \{s\})$. If $\shat = s^*$, then we're done.
Otherwise, $\shat < s^*$.
Let $\sigma: S \setminus \{\shat\} \to S \setminus \{s^*\}$ be a function where
$\sigma(s^*) = \shat$ and $\sigma(s) = s$ for all $s \in S \setminus \{s^*, \shat\}$.
Then by \cref{thm:cancelable-agg}, we get
$u(S \setminus \{s^*\}) \ge u(S \setminus \{\shat\}) = \max_{s \in S} u(S \setminus \{s\})$.
\end{proof}

\subsection{EEFX for Cancelable Valuations}
\label{sec:cancelable:eefx}

\citet{barman2020approximation} proved that the Envy-Cycle-Elimination (ECE) algorithm
outputs an \efxZero\ allocation for goods with ordered additive valuations.
Their proof can be easily adapted to ordered cancelable valuations using \cref{thm:cancelable-last}.
See Appendix~\ref{sec:cancelable-extra:ece} for details.

The output of \cref{alg:eefx} (c.f.~\cref{sec:eefx}) is \eefxZero\ for goods having cancelable valuations.
The proof is similar to that of \cref{lem:eefx}.
The key ingredients in the proof are \cref{lem:properties-of-bijection}, \cref{thm:cancelable-agg},
and ECE's output being \efxZero\ for ordered cancelable valuations.
See Appendix~\ref{sec:cancelable-extra:eefx} for details.

For chores too, an \eefxZero\ allocation can be computed in polynomial time
by adapting the techniques of \cref{sec:chores} to cancelable valuations using \cref{sec:cancelable:ordered}.
See Appendix~\ref{sec:cancelable-extra:chores} for details.

\section{Discussion}
\label{sec:discussion}

We have presented epistemic \efx\ and minimum \efx\ share, two new fairness concepts which are defined using the well-known \efx\ fairness notion.

Our work reveals many open problems. We have already mentioned that, in addition to being \eefx\ and \mxs, the allocations computed by \cref{alg:eefx} are $2/3$-\mms\ as well~\citep{barman2020approximation}. Are there \eefx\ allocations with better \mms\ guarantees possible? Can they be computed in polynomial time? Combining \eefx\ with other important fairness notions, like \ef1, is another interesting research direction.

Furthermore, proving that \eefx\ and \mxs\ are compatible with efficiency could nicely complement their fairness properties.
For example, one question we have left open is whether Pareto-optimal \eefx\ allocations exist. Also, notice that the allocation we showed to be \eefx\ in \cref{example:firstExample} has maximum {\em social welfare} (i.e., maximum total value for the agents). It is tempting to conjecture that this is the case in general and \eefx\ has a low {\em price of fairness}~\citep{CKKK12,BFT11}. This deserves investigation for both \eefx\ and \mxs.

Finally, it is certainly interesting to explore whether the results we present here can be extended beyond additive/cancelable valuations.
Recently, \cite{AN24} have proved the existence of $\eefx$ allocations for \emph{monotone} valuations. They further show that even for an arbitrary number of (more than one) agents with identical submodular valuations, it is $\pls$-hard to compute $\eefx$ allocations and it requires exponentially-many value queries to do so.

\appendix

\section{Defining EFX for Non-Additive Valuations}
\label{sec:efx-defn}

We have adopted the original definition of \efx\ by \citet{CKMPSW19} in this paper.
However, \citet{plaut2020almost} used a simpler definition that does not have the restriction $v_i(g)>0$ (see \cref{defn:efx}) and yields a slightly stronger fairness notion, called \efxZero~\citep{KSV20}.

\defnEfxZero*

Using \efxZero, we can define \eefxZero\ and \mxs$_0$ in a similar way we defined \eefx\ and \mxs.

The difference between \efx\ and \efxZero\ holds significance in our work:
An MMS allocation is always \eefx\ (\cref{thm:mms_eefx}), but may not be \eefxZero.

\begin{example}
Consider the instance with two agents having identical valuations $1$ and $0$ for two items $a$ and $b$, respectively.
Note that the \mms\ share of both agents is zero and, hence, the allocation in which agent 1 gets both items is \mms.
However, no \eefxZero\ certificate exists for agent 2 with the empty bundle.
\end{example}

Moreover, while \efxZero\ can be trivially extended to the setting where agents may have
non-additive (monotone) valuations, it is not obvious how to extend \efx\ to this setting.
E.g., in an allocation $X$, agent $i$ may envy agent $j$ even if
$v_i(g) = 0$ for every $g \in X_j$ (this necessarily requires non-additive valuations).
In this section, we propose an extension of \efx\ to general monotone valuations.
For any two sets $S$ and $T$ of items, define $v_i(S \mid T) \defeq v_i(S \cup T) - v_i(T)$.

\begin{definition}[\efx]
An agent $i$ is \emph{EFX-satisfied} by allocation $X = (X_1, \ldots, X_n)$ if for every other agent $k$,
we have $v_i(X_i) \ge v_i(X_k)$ or
$v_i(X_i) \ge \max\{v_i(X_j \setminus S): S \subseteq X_j \textrm{ and } v_i(S \mid X_i) > 0\}$.
The allocation $X$ is \emph{EFX} if every agent is EFX-satisfied by $X$.
\end{definition}

\begin{claim}
Consider a fair division instance with general monotone valuation functions.
In an allocation $X$, if an agent is \efxZero-satisfied, then she is also \efx-satisfied.
\end{claim}

\begin{claim}
Consider a fair division instance with general monotone valuation functions
and positive marginals (i.e., $v_i(g \mid S) > 0$ for every $i \in [n]$, every $S \subseteq [m]$,
and every $g \in [m] \setminus S$). Then in every allocation $X$,
an agent is \efxZero-satisfied iff she is \efx-satisfied.
\end{claim}

One can show that \cref{thm:mms_eefx} (MMS $\Rightarrow$ EEFX)
also works for general monotone valuations.

\section{Details on Cancelable Valuations}
\label{sec:cancelable-extra}

Here we give details that were omitted from \cref{sec:cancelable}.

\subsection{Equivalence of Definitions}
\label{sec:cancelable-extra:defn-equiv}

\thmCancDefnEquiv*
\begin{proof}
Let $P(S_1, S_2, T)$ be the predicate $u(S_1) \ge u(S_2) \Rightarrow u(S_1 \cup T) \ge u(S_2 \cup T)$.
Then the contrapositive of $P(S_2, S_1, \{g\})$ is equivalent to \cref{defn:cancelable}.
Hence, if $P(S_1, S_2, T)$ holds for all $T \subseteq [m]$ and $S_1, S_2 \subseteq [m] \setminus T$,
then $u$ is cancelable.

Now suppose $u$ is cancelable. Let $T \defeq \{t_1, \ldots, t_k\} \subseteq [m]$
and $S_1, S_2 \subseteq [m] \setminus T$.
For any $i \in \{0\} \cup [k]$, let $T_i \defeq \{t_1, \ldots, t_i\}$.
Then $u(S_1 \cup T_i) > u(S_2 \cup T_i) \Rightarrow u(S_1 \cup T_{i-1}) > u(S_1 \cup T_{i-1})$
for all $i \in [k]$, by \cref{defn:cancelable}.
Hence, $u(S_1 \cup T) > u(S_2 \cup T) \Rightarrow u(S_1) > u(S_2)$,
which is equivalent to the contrapositive of $P(S_2, S_1, T)$.
Hence, $P(S_1, S_2, T)$ holds for all $T \subseteq [m]$ and $S_1, S_2 \subseteq [m]$
if $u$ is cancelable.
\end{proof}

\subsection{The Envy Cycle Elimination Algorithm}
\label{sec:cancelable-extra:ece}

The Envy Cycle Elimination algorithm by \citet{lipton2004approximately}
outputs an \efxZero\ allocation when agents have ordered additive valuations
\citep{plaut2020almost,barman2020approximation}.
We show that this holds even for ordered cancelable valuations.

We first recall the Envy Cycle Elimination algorithm for goods.
For an allocation $X$, let $E_X$ be a directed graph (called the \emph{envy graph})
where the agents are vertices and there is an edge from agent $i$ to $j$
if $i$ envies $j$, i.e., $v_i(X_i) < v_i(X_j)$.
A cycle in $E_X$ is called an \emph{envy cycle} for $X$.
Formally, let $C \defeq (i_1, i_2, \ldots, i_k)$ and let $i_0 \defeq i_k$.
Then $C$ is called an \emph{envy cycle} in $X$ if
agent $i_{t-1}$ envies agent $i_t$ in $X$ for each $t \in [k]$.
\emph{Resolving} the envy cycle $C$ is an operation where agent $i_{t-1}$ gets
agent $i_t$'s bundle for each $t \in [k]$. Formally, the allocation
$Y \defeq \resolve(X, C)$ is defined as
\[ Y_j \defeq \begin{cases}
X_{i_t} & \textrm{ if } j = i_{t-1}, \textrm{ for some } t \in [k]
\\ X_j & j \not\in C
\end{cases}. \]
If $E_X$ is acyclic, it must have a vertex of in-degree 0, called an \emph{unenvied agent}.

In the Envy Cycle Elimination algorithm, we repeatedly resolve envy cycles
till the envy graph becomes acyclic, and then give a good to an unenvied agent.
See \cref{algo:ece-goods} for a more precise description.

\begin{algorithm}[htb]
\caption{$\celim$}
\label{algo:ece-goods}
\begin{algorithmic}[1]
\Require A fair division instance $\instance$.
\Ensure An allocation $X$ for $\instance$.
\State $X_i \gets \emptyset$ for each $i \in [n]$.
\For{$g$ from $1$ to $m$}
    \While{$X$ contains an envy cycle $C$}
        \State\label{alg-line:ece-goods:resolve}$X \gets \resolve(X, C)$.
    \EndWhile
    \State Pick an arbitrary unenvied agent $i$.
    \State \label{alg-line:ece-goods:alloc}$X_i \gets X_i \cup \{g\}$
\EndFor
\end{algorithmic}
\end{algorithm}

\begin{lemma}
\Cref{algo:ece-goods} always terminates, and runs in polynomial time,
given oracle access to players' valuations for any subset of goods.
\end{lemma}
\begin{proof}
Follows from the proof of Theorem~2.1 in \citet{lipton2004approximately}.
\end{proof}

\begin{lemma}
\label{thm:ece-goods-resolve-efx}
Let $X$ be an \efxZero\ allocation of goods.
Then $Y \defeq \resolve(X, C)$ is also \efxZero, where $C$ is an envy cycle for $X$.
\end{lemma}
\begin{proof}[Proof sketch]
Follows from the proof of Theorem~2.1 in \citet{lipton2004approximately}.
The key idea is that $\resolve$ doesn't change the bundles, only changes their owners,
and agents' envy is essentially towards bundles, regardless of who owns them.
\end{proof}

\begin{lemma}
\label{thm:ece-efx-canc}
Let $\instance$ be a fair division instance of goods with ordered cancelable valuations.
Then the output of $\celim(\instance)$ (envy-cycle-elimination) is \efxZero\ for $\instance$.
\end{lemma}
\begin{proof}
Let $X^{(t)}$ be the allocation $X$ in \cref{algo:ece-goods}
immediately after $t$ iterations of the for loop.
We will show that $X^{(t)}$ is \efxZero\ for every $t \in \{0\} \cup [m]$ using induction.

Initially, $X^{(0)}_j = \emptyset$ for each $j \in [n]$, so $X^{(0)}$ is \efxZero.
Now let $t \in [m]$ and assume $X^{(t-1)}$ is \efxZero.
In the $t\Th$ iteration, let $Y$ be the allocation immediately after all
envy cycles have been resolved, and let $i$ be the unenvied agent selected to receive good $t$.
Since $X^{(t-1)}$ is \efxZero, $Y$ is also \efxZero\ by \cref{thm:ece-goods-resolve-efx}.
$X^{(t)}_i = Y_i \cup \{t\}$ and $X^{(t)}_j = Y_j$ for all $j \in [n] \setminus \{t\}$.

Suppose $X^{(t)}$ is not \efxZero. Then there exists an agent $j \in [n]$ that
\efxZero-envies some agent $k$ in $X^{(t)}$, i.e.,
$v_j(X^{(t)}_j) < \max_{g \in X^{(t)}_k} v_j(X^{(t)}_k \setminus \{g\})$.

Agent $i$ receives a good and the other agents don't, so she remains \efxZero-satisfied, so $j \neq i$.
Also, $k = i$, because otherwise $j$ and $k$ would have the same bundles in $X^{(t)}$ and $Y$.
By \cref{thm:cancelable-last}, we get
\[ v_j(Y_j) = v_j(X^{(t)}_j) < \max_{g \in X^{(t)}_k} v_j(X^{(t)}_i \setminus \{g\})
= v_j(X^{(t)}_i \setminus \{t\}) = v_j(Y_i). \]
This is a contradiction, since $i$ is unenvied in $Y$.
Hence, $X^{(t)}$ is \efxZero.

By mathematical induction, we get that $X^{(t)}$ is \efxZero\ for all $t \in [m]$.
Hence, $X^{(m)}$, which is the output of \cref{algo:ece-goods}, is also \efxZero.
\end{proof}

\subsection{Computing EEFX Allocations}
\label{sec:cancelable-extra:eefx}

We now show that \cref{alg:eefx} (c.f.~\cref{sec:eefx}) outputs an \eefxZero\ allocation when agents have cancelable valuations.

\begin{lemma}
\label{thm:canc-implies-ord-canc}
In a fair division instance, if an agent $i$'s valuation function $v_i$ is cancelable,
then her ordered valuation function $\vord_i$ (c.f.~\cref{sec:prelims:order}) is also cancelable.
\end{lemma}
\begin{proof}
For any $S \subseteq [m]$, we have $\vord_i(S) \defeq v_i(r_i^{-1}(S))$.
Let $j \in [m]$ and $S_1, S_2 \subseteq [m] \setminus \{j\}$. Let $r_i^{-1}(j) = g$.
Then by cancelability of of $v_i$, we get
\begin{align*}
& \vord_i(S_1 \cup \{j\}) > \vord_i(S_2 \cup \{j\})
\Leftrightarrow v_i(r_i^{-1}(S_1) \cup \{g\}) > v_i(r_i^{-1}(S_2) \cup \{g\})
\\ &\Rightarrow v_i(r_i^{-1}(S_1)) > v_i(r_i^{-1}(S_2))
\Leftrightarrow \vord_i(S_1) > \vord_i(S_2)
\end{align*}
Hence, $\vord_i$ is also cancelable.
\end{proof}

\begin{lemma}
\label{thm:eefx-canc}
For a fair division instance of goods with cancelable valuations,
the output of \cref{alg:eefx} (c.f.~\cref{sec:eefx}) is \eefxZero.
\end{lemma}
\begin{proof}
Consider agent $i\in [n]$ and let $\pi_i$ be the bijection defined in \cref{lem:properties-of-bijection}. Define the allocation $Y$ with $Y_j = \{\pi_i(g): g \in \Xord_j\}$ for $j \in [n]$. Since $\pi_i$ is a bijection, allocation $Y$ is well-defined. Also, by \cref{lem:properties-of-bijection}, $Y_i = X_i$. We will prove that $Y$ is an \eefxZero\ certificate for agent $i$ with bundle $X_i$.

By \cref{thm:canc-implies-ord-canc}, we get that $\vord_i$ is ordered and cancelable for all $i \in [n]$.
Applying \cref{thm:cancelable-agg} with $S = r_i(Y_i)$, $T = \Xord_i$, and $\sigma = r_i\cdot\pi_i$, we get
$v_i(Y_i) = \vord_i(r_i(Y_i)) \ge \vord_i(\Xord_i)$.
For any $j \neq i$, applying \cref{thm:cancelable-agg} with $S = \Xord_j$, $T = r_i(Y_j)$,
and $\sigma = (r_i\cdot\pi_i)^{-1}$, we get
\[ \max_{\gord \in \Xord_j} \vord_i(\Xord_j \setminus \{\gord\})
    \ge \max_{h \in r_i(Y_j)} \vord_i(r_i(Y_j) \setminus \{h\})
    = \max_{g \in Y_j} v_i(Y_j \setminus \{g\}). \]
By \cref{thm:ece-efx-canc}, $\Xord$ is \efxZero\ for $\ordInst$. Hence, for any $j \neq i$, we get
\[ v_i(Y_i) \ge \vord_i(\Xord_i) \ge \max_{\gord \in \Xord_j} \vord_i(\Xord_j \setminus \{\gord\})
    \ge \max_{g \in Y_j} v_i(Y_j \setminus \{g\}). \]
Hence, agent $i$ is \efxZero-satisfied by $Y$, so $Y$ is $i$'s \eefxZero-certificate for $X$.
\end{proof}

\subsection{EEFX for Chores}
\label{sec:cancelable-extra:chores}

We show that the modification of \cref{alg:eefx} for chores (c.f.~\cref{sec:chores})
outputs an \eefxZero\ allocation when agents have cancelable valuations.
First, let's formally define \efxZero. We can define \eefxZero\ and \mxs$_0$ for chores
similarly to how we defined \eefx\ and \mxs\ for chores.

\begin{definition}[\efxZero]
An agent $i$ is \efxZero\emph{-satisfied} by an allocation $X = (X_1, \ldots, X_n)$ of chores
if for every other agent $j$, we have
$\displaystyle \max_{c \in X_i} |v_i(X_i \setminus \{c\})| \le |v_i(X_j)|$.
The allocation $X$ is \efxZero\ if every agent is \efxZero-satisfied by $X$.
\end{definition}

As discussed in \cref{sec:chores}, the envy-cycle-elimination algorithm must be modified to work for chores.
We now formally define the top-trading envy graph,
and then formally describe the modified algorithm in \cref{algo:ece-chores}.
These modifications are based on Section~5 of \citet{bhaskar2021}.

For an allocation $X$, we say that agent $i$ top-envies agent $j$ if
$|v_i(X_i)| \ge |v_i(X_j)|$ and $j \in \argmin_{k \in [n] \setminus \{i\}} |v_i(X_k)|$.
Let $T_X$ be a directed graph (called the \emph{top-trading envy graph}) where
the agents are vertices and there is an edge from agent $i$ to $j$ if $i$ top-envies $j$.
A cycle in $T_X$ is called a \emph{top-trading envy cycle} for $X$.
Formally, let $C \defeq (i_1, i_2, \ldots, i_k)$ and let $i_0 \defeq i_k$.
Then $C$ is called a \emph{top-trading envy cycle} in $X$ if
agent $i_{t-1}$ top-envies agent $i_t$ in $X$ for each $t \in [k]$.
\emph{Resolving} the cycle $C$ is an operation where agent $i_{t-1}$ gets
agent $i_t$'s bundle for each $t \in [k]$. Formally, the allocation
$Y \defeq \resolve(X, C)$ is defined as
\[ Y_j \defeq \begin{cases}
X_{i_t} & \textrm{ if } j = i_{t-1}, \textrm{ for some } t \in [k]
\\ X_j & j \not\in C
\end{cases}. \]
If $T_X$ is acyclic, it must have a vertex of out-degree 0.
Such an agent is \emph{unenvious}, i.e., doesn't envy any other agent.

\begin{algorithm}[htb]
\caption{$\ttcelim$}
\label{algo:ece-chores}
\begin{algorithmic}[1]
\Require A fair division instance $\instance$ of chores.
\Ensure An allocation $X$ for $\instance$.
\State $X_i \gets \emptyset$ for each $i \in [n]$.
\For{$c$ from $m$ to $1$}
    \While{$X$ contains a top-trading envy cycle $C$}
        \State\label{alg-line:ece-chores:resolve}$X \gets \resolve(X, C)$.
    \EndWhile
    \State Pick an arbitrary unenvious agent $i$.
    \State \label{alg-line:ece-chores:alloc}$X_i \gets X_i \cup \{c\}$
\EndFor
\end{algorithmic}
\end{algorithm}

\begin{lemma}
\Cref{algo:ece-chores} always terminates, and runs in polynomial time,
given oracle access to players' valuations for any subset of chores.
\end{lemma}
\begin{proof}
Follows from Section~5 of \citet{bhaskar2021}.
\end{proof}

\begin{lemma}
\label{thm:ece-efx-canc-chores}
Let $\instance$ be a fair division instance of chores with ordered cancelable valuations.
Then the output of $\ttcelim$ is \efxZero\ for $\instance$.
\end{lemma}
\begin{proof}
It easily follows using ideas from
Section~5 of \citet{bhaskar2021} and \cref{thm:ece-efx-canc}.
\end{proof}

\begin{lemma}
\label{thm:eefx-canc-chores}
For a fair division instance of chores with cancelable valuations, the output of
modified \cref{alg:eefx} (i.e., using $\ttcelim$ instead of $\celim$) is \eefxZero.
\end{lemma}
\begin{proof}
Consider agent $i\in [n]$ and let $\pi_i$ be the bijection defined in \cref{lem:properties-of-bijection}. Define the allocation $Y$ with $Y_j = \{\pi_i(c): c \in \Xord_j\}$ for $j \in [n]$. Since $\pi_i$ is a bijection, allocation $Y$ is well-defined. Also, by \cref{lem:properties-of-bijection}, $Y_i = X_i$.
We will show that agent $i$ is \efxZero-satisfied by $Y$.
That would prove that $Y$ is an \eefxZero\ certificate for agent $i$ with bundle $X_i$.

By \cref{thm:canc-implies-ord-canc}, we get that $\vord_i$ is ordered and cancelable for all $i \in [n]$.
Applying \cref{thm:cancelable-agg} with $S = r_i(Y_i)$, $T = \Xord_i$,
and $\sigma = r_i\cdot\pi_i$, we get
\[ \min_{c \in Y_i} v_i(Y_i \setminus \{c\})
    = \min_{h \in r_i(Y_i)} \vord_i(r_i(Y_i) \setminus \{h\})
    \ge \min_{\cord \in \Xord_i} \vord_i(\Xord_i \setminus \{\cord\}). \]
Applying \cref{thm:cancelable-agg} with $S = \Xhat_j$, $T = r_i(Y_j)$,
and $\sigma = (r_i\cdot\pi_i)^{-1}$, we get $\vord_i(\Xord_j) \ge \vord_i(r_i(Y_j)) = v_i(Y_j)$.
By \cref{thm:ece-efx-canc}, $\Xord$ is \efxZero\ for $\ordInst$. Hence, for any $j \neq i$, we get
\[ \min_{c \in Y_i} v_i(Y_i \setminus \{c\})
    \ge \min_{\cord \in \Xord_i} \vord_i(\Xord_i \setminus \{\cord\})
    \ge \vord_i(\Xord_j) \ge v_i(Y_j). \]
Hence, agent $i$ is \efxZero-satisfied by $Y$, so $Y$ is $i$'s \eefxZero-certificate for $X$.
\end{proof}

\end{document}